\documentclass{ws-rv961x669}
\usepackage{ws-rv-van}
\usepackage{ws-rv-thm}

\makeindex
\begin{document}

\title{Neutrino Physics and Astrophysics}

\chapter[IMB, Kamiokande and Super-K]{Neutrino Astronomy with IMB, Kamiokande and Super Kamiokande\footnotemark}\footnotetext{To be published in Neutrino Physics and
Astrophysics, edited by F. W. Stecker, in the Encyclopedia of Cosmology II,
edited by G. G. Fazio, World Scientific Publishing Company, Singapore, 2022.}
\author[J. LoSecco]{John M. LoSecco}
\address{The University of Notre Dame du Lac}
\begin{abstract}
  Some of the earliest work on neutrino astronomy was accomplished by
  a class of underground detectors primarily designed for particle physics goals.
  These detectors used inexpensive water to obtain the large masses needed
  to observe the very low interaction rates expected from neutrinos.
  They exploited the relatively large light attenuation length and the index
  of refraction of the water to get a very inexpensive cost per thousand
  tons of detector.

  The results obtained from these pioneering neutrino detectors have included real time observation of solar neutrinos, supernova neutrinos, and atmospheric neutrinos.  Searches for neutrino point sources, dark matter and primordial magnetic monopoles were also made
using them.
\end{abstract}
\newpage
\tableofcontents
\newpage
\section{Introduction}
Neutrino astrophysics is now a well established scientific discipline.
Neutrinos have been observed from a broad selection of natural sources,
including the sun, supernovae, cosmic ray interactions in the atmosphere
and astrophysical sources.  Future possibilities include relic supernova neutrinos,
neutrinos from the Big Bang, and perhaps neutrinos from dark matter annihilation.  Neutrinos have even left a footprint on the structure of the
cosmic microwave background radiation.

The growth of neutrino astrophysics to maturity has taken decades of work.  For the most part this is because theoretical neutrino astronomy was well ahead of the technology.  The goal of using neutrinos to study and understand astrophysical processes required
a deeper understanding of how neutrinos are classified and how they propagate.
The class of detectors discussed here has
played key roles in neutrino astrophysics discoveries.
\section{History}
\subsection{Requirements}
Neutrinos have no electromagnetic or strong nuclear interactions.  They are
normally observed via the weak interaction.  A great deal can be inferred from observations of the final state of such neutrino interactions. They inform about the direction, energy and the type of neutrinos involved. 

There are three known kinds of neutrinos, called "flavors".  Neutrinos can also
interact via the two different manifestations of the weak interaction.  The
dominant charged current interaction converts the neutrino into a charged
particle that can be observed.  The neutral current weak interaction has both
a neutrino entering and leaving the reaction which makes it harder to
analyze.
\begin{figure}[h]
	\centering
	\includegraphics[width=0.7\linewidth]{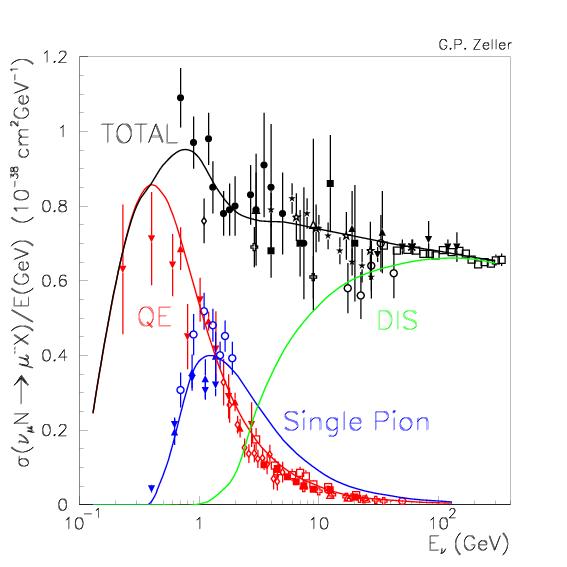}
	\caption{The neutrino nucleon cross section on an isoscalar target divided by the neutrino energy in GeV. Figure from \cite{Fo+Ze}}
	\label{fig:zellertotalnumode}
\end{figure}
Figure \ref{fig:zellertotalnumode} makes it clear why neutrinos are hard to observe.
The vertical scale displays the neutrino nucleon cross section divided by the
neutrino energy (in GeV) as a function of the neutrino energy.
The horizontal axis is a log scale running from 0.1 to 100 GeV.
The cross section never gets above about $\sigma$ = 10$^{-38}$ cm$^{2}$ GeV$^{-1}$ per neutrino
per nucleon (proton or neutron) at 1 GeV.  Water has a density of about 1000 kg m$^{-3}$,
ie, about $\rho = 6.02 \times 10^{23}$ nucleons per cubic centimeter. This
gives a scattering length for a 1 GeV neutrino in water at about
$(\rho \sigma)^{-1} = 1.7 \times 10^{14}$ cm.
However, because of the energy dependence of the neutrino cross section, many neutrinos of interest to astrophysics that have much lower energies, have 
much lower interaction probabilities.

The rate of observed neutrino interactions depends on several factors.
As mentioned above, the cross section is an increasing function of the neutrino
energy.  The interaction event rate also depends on how many neutrinos traverse the detector per unit time and the number of targets, electrons or nucleons, in the detector.  The goal here is to observe neutrinos in nature so there is no way
to control the energy distribution or the flux of neutrinos so most detector designs
concentrate on building the most massive detector possible and maximizing the
detection efficiency.  Sometimes the choice of the target material can
make a difference with a higher cross section for some neutrino reactions.

Beyond direct detection of the neutrino interaction there are other methods
that have been used to increase the rate of observed neutrino interactions
by looking for indirect evidence.  For example, upward going muons are mostly
created by neutrino interactions in the Earth below that then penetrate
a substantial distance in the rock.  Thus, the indirect detection method
has extended the effective mass of the neutrino target, this at the cost of missing some detailed information about the signal.  However, one can achieve a much more massive
target with a much smaller detector.

\subsubsection{Neutrino Electron Scattering}
$\nu_{e} + e \rightarrow \nu_{e} + e$ and the similar reactions
$\nu_{\mu,\tau} + e \rightarrow \nu_{\mu\tau} + e$ are very well understood since they only involve leptons\cite{nue}.  In most cases the neutrino energy is much
higher than the electron mass so the electron direction is close to the neutrino direction and the reaction can be used for pointing.

Defining $y$ to be the fractional energy loss of the neutrino,
$y = E_{e}/E_{\nu}$, in the small angle approximation one finds

\begin{equation}
E_{e} \theta_{e}^{2} = 2 m_{e} (1-y),
\end{equation}
Therefore, since $0 \le y \le 1$, it follows that $E_{e} \theta_{e}^{2} < 2 m_{e}$.
\begin{figure}
	\centering
	\includegraphics[width=0.8\linewidth]{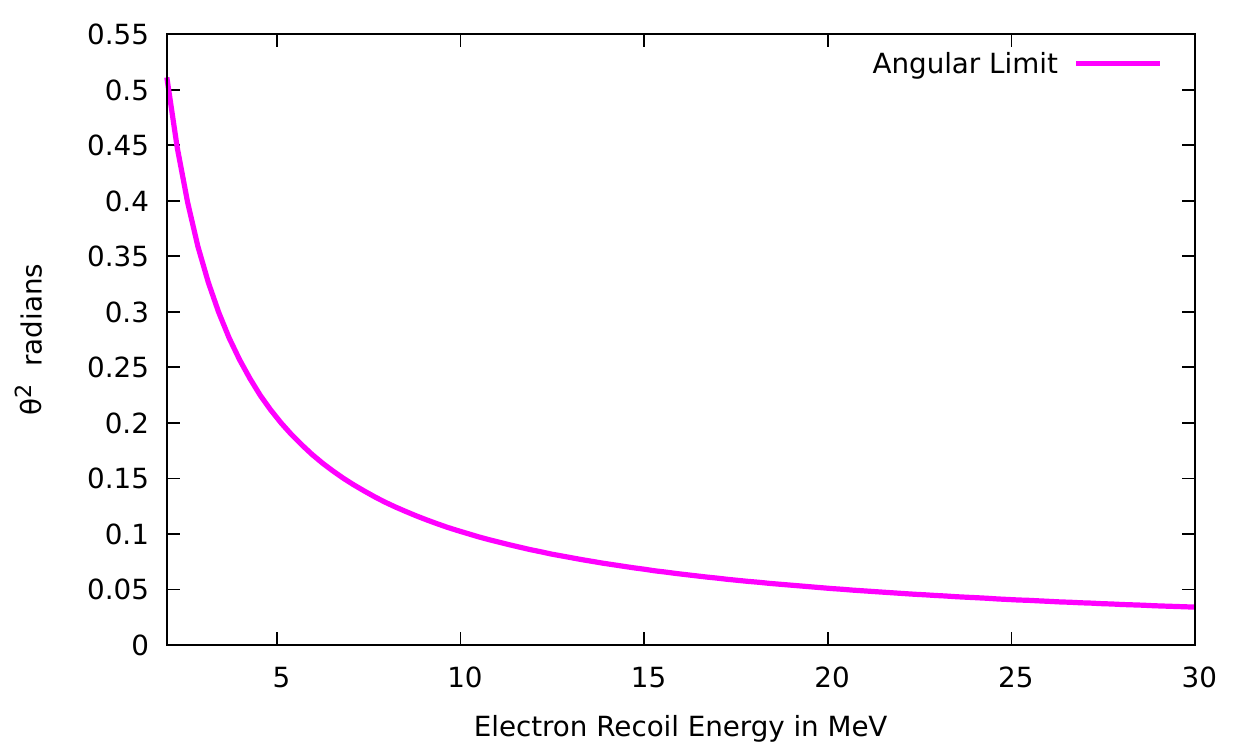}
	\caption{The upper limit on the angular distribution for $\nu + e^{-}$ scattering as a function of the electron recoil energy}
	\label{fig:nueangle}
\end{figure}

This reaction is very useful for astrophysics since the recoil electron
points back to the neutrino source, (See Figure \ref{fig:nueangle}).  Also, most astrophysical sources
will have a strong electron neutrino component at Earth
(See sections 4 and 5.3).

The neutrino cross section\cite{nue} is very small, about $10^{-45}$ cm$^{2}$ at 1 MeV for $\nu_{\mu}$ or $\nu_{\tau}$ and about $10^{-44}$ cm$^{2}$ at 1 MeV for
$\nu_{e}$. the differential cross section,
\begin{equation}
\frac{d\sigma^{\nu_{\mu},\bar{\nu}_{\mu}}}{dy}=\frac{2G_{F}^{2}m_{e}}{\pi}E_{\nu} g_{L,R}^{2} +    g_{R,L}^{2} (1-y^2) -g_{L}g_{R} \frac{M_{e}y}{E_{\nu}},
\end{equation}
where $g_{L}$ and $g_{R}$ are the couplings to left and right handed electrons.
$2g_{L}=g_{V}+g_{A}$ and $2g_{R}=g_{V}-g_{A}$.  $g_{V}$ and $g_{A}$
are the vector and axial vector couplings.  There are substantial radiative
corrections to this simple result.  For $\nu_{e}$ scattering there is a charged current contribution which requires replacing $g_{L}$ with
$g_{L}+1$.


\begin{figure}[h]
	\centering
	\includegraphics[width=0.75\linewidth]{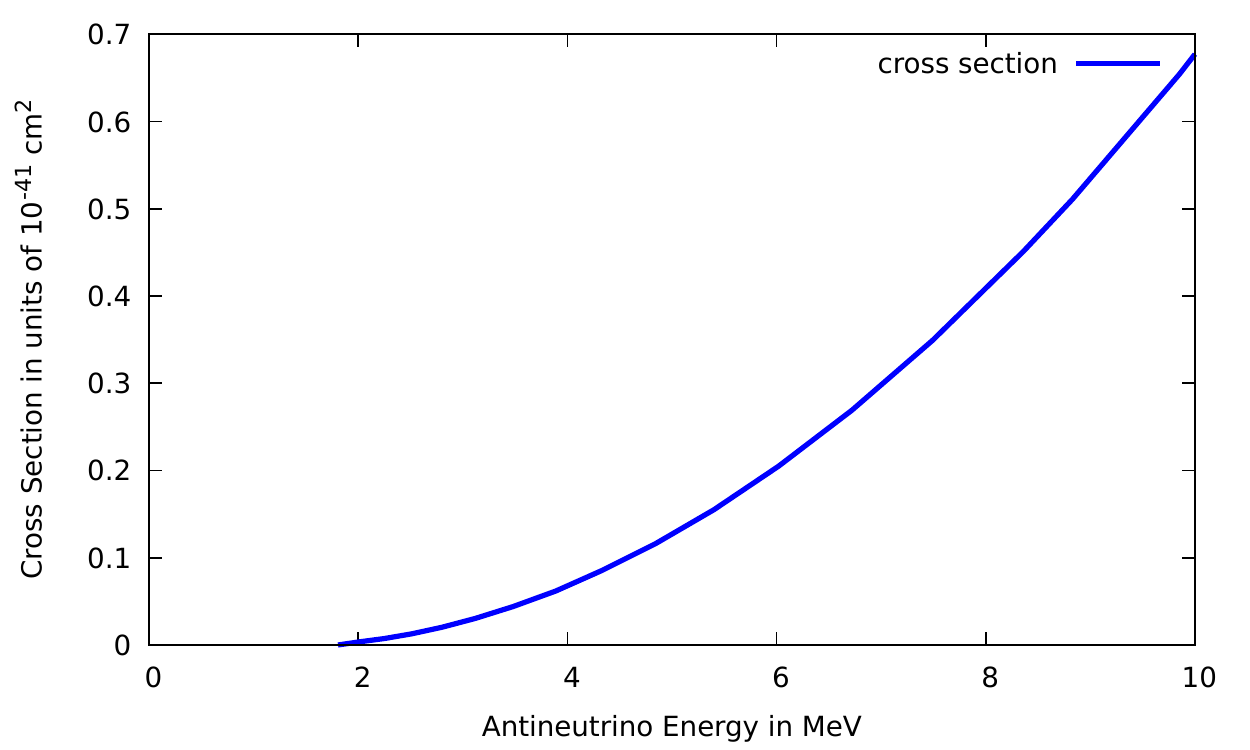}
	\caption{The $\bar{\nu}_e + p \rightarrow e^{+} + n$ cross section
		in units of $10^{-41} cm^{2}$ for nuclear reactor energy antineutrinos.}
	\label{fig:reactorxsec}
\end{figure}
\begin{figure}[h]
	\centering
\includegraphics[width=0.75\linewidth]{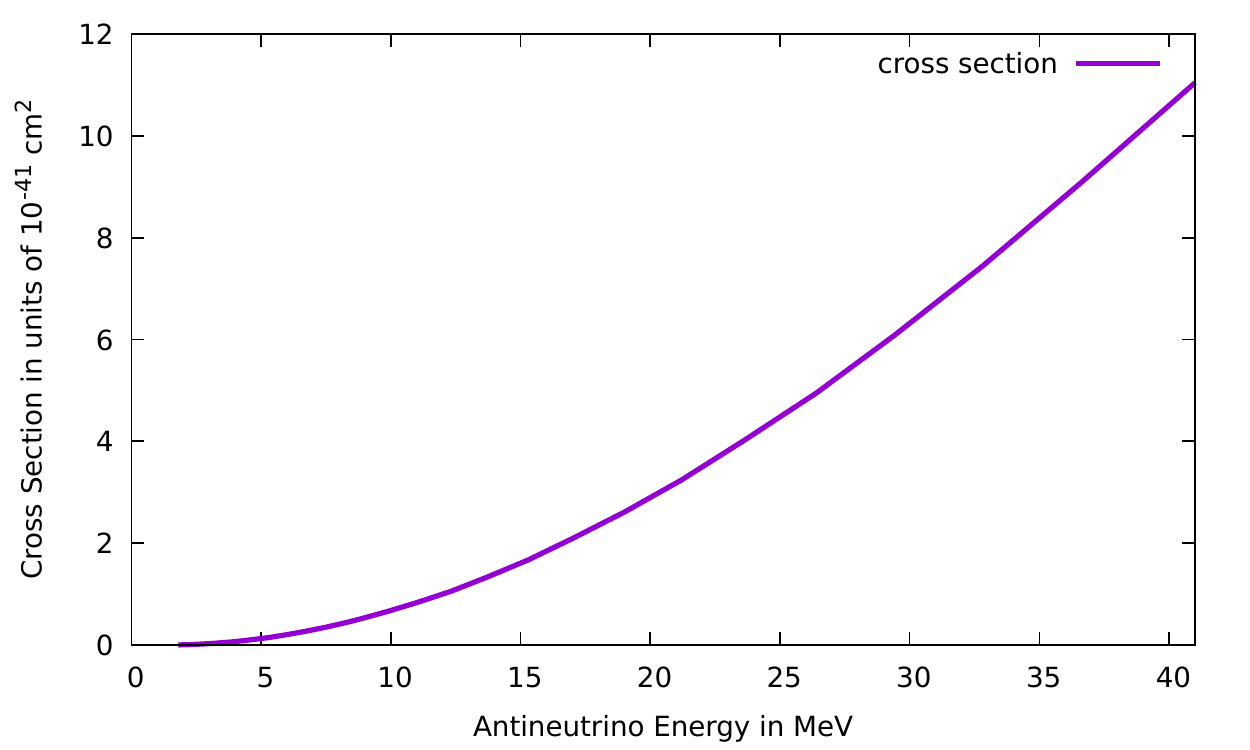}
	\caption{The $\bar{\nu}_e + p \rightarrow e^{+} + n$ cross section
	in units of $10^{-41} cm^{2}$ for supernova range energies.}
\label{fig:reactorxsecSN}
\end{figure}
A special case of neutrino electron scattering occurs at very high energies
where the center of mass energy is close to the mass of the $W^{-}$ weak
boson\cite{GlasRes}.  The threshold for this reaction is 6.3 PeV.  The cross
section is expected to increase by about a factor of 200 near this resonance.
The final state, the decay products of $W^{-}$ decay is not the simple
two body final state of low energy $\nu_{e} + e$ scattering.
\subsubsection{Antineutrino Proton Scattering}
\begin{figure}[h]
	\includegraphics[width=0.496\linewidth]{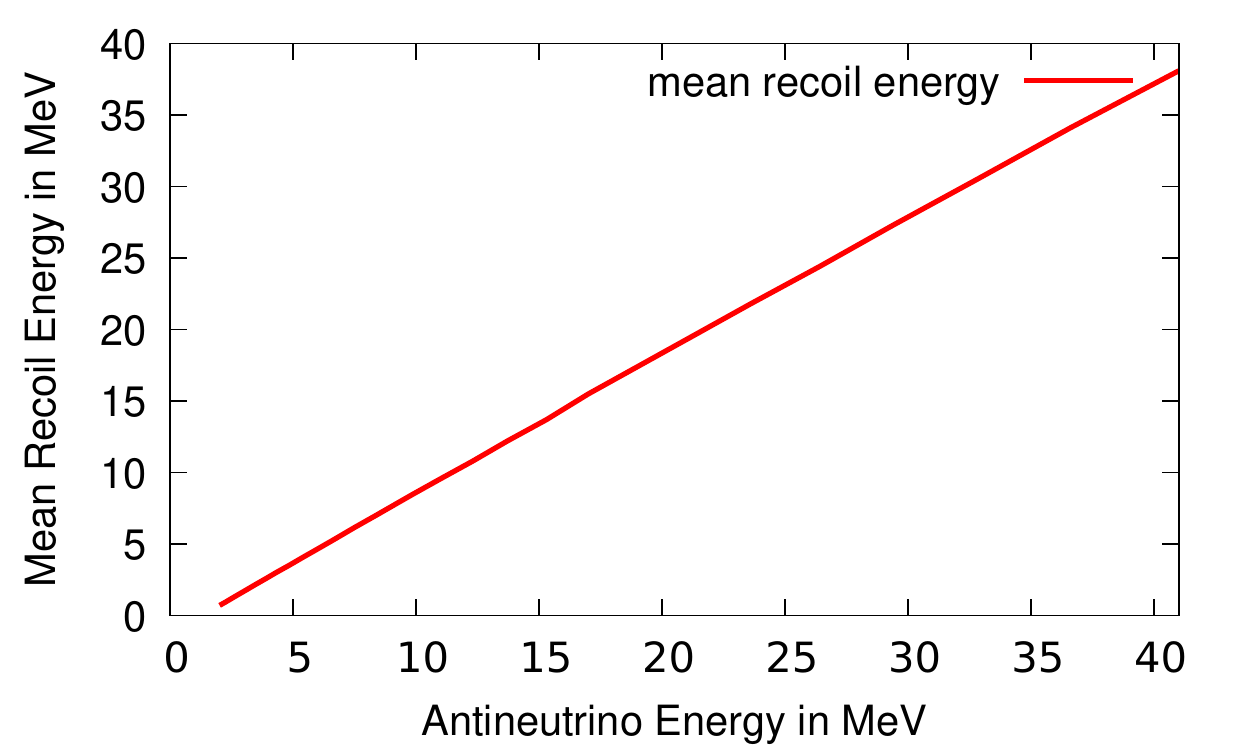}
	\includegraphics[width=0.496\linewidth]{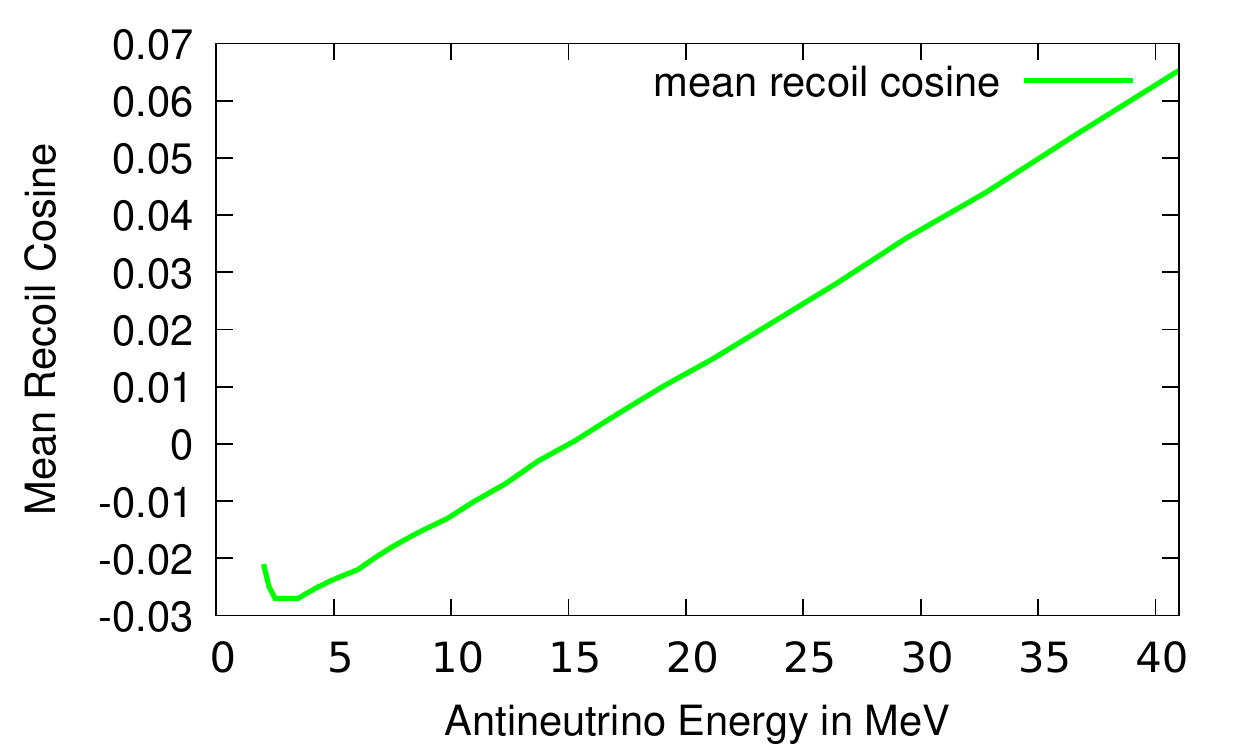}
	\caption{The $\bar{\nu}_e + p \rightarrow e^{+} + n$ positron recoil
		energy in MeV showing good fidelity with the neutrino energy (left).
		The mean cosine of the recoil positron is shown (right).
		The positron does not provide useful directional information being nearly
		isotropic at these energies.}
	\label{fig:reactorAngE}
\end{figure}
The reaction
\begin{equation}
 \bar{\nu}_e + p \rightarrow e^{+} + n,
 \label{IBD} 
\end{equation}
the charged current scattering of an electron antineutrino from a proton has been a very effective tool in neutrino physics\cite{nubarp}.  This reaction was used as a basis for the discovery of the neutrino. 

The presence of the $e^{+}$ and $n$ in the final state allows the use of coincidence methods of finding both in order to enhance the true neutrino
signals while rejecting backgrounds that would make a single $e^{+}$ or $n$.

The process (\ref{IBD}) is known as Inverse Beta Decay, IBD.
The cross section for IBD 
has an approximately quadratic dependence on the neutrino energy,
(See Figures \ref{fig:reactorxsec} and \ref{fig:reactorxsecSN}).
For supernova and reactor antineutrinos this reaction provides good energy
resolution since almost all the energy is carried off by the $e^{+}$
and can be reconstructed (See Figure \ref{fig:reactorAngE}).

The recoiling positron lacks directionality (See Figure \ref{fig:reactorAngE}), so it is
not useful
in locating a signal if it is the only source of information about the source.
Since the neutrino momentum is conserved, the forward going neutron
carries this directional information.
Unfortunately the neutron has little kinetic energy and the initial direction
is not practically detectable.

\subsubsection{High Energy Neutrino Interactions}
Neutrinos have two properties that make them very appealing for astrophysics.
They are electrically neutral so their point of origin is not distorted from
their source to a detection.  Neutrinos are also very penetrating so they
carry information directly from the interior of celestial objects that
photons can not provide.  This later feature makes them difficult to observe
but massive detectors and large fluxes compensate for this weakness.

While most neutrinos occurring from low energy astrophysical processes are
of the electron type, since they basically involve converting protons into neutrons, at higher energies the strong interactions provide an efficient method to make muon type neutrinos.  The lightest strongly interacting
particle, the pion, decays 99.988\% of the time into a muon neutrino.
Any pion not absorbed will make a neutrino.

One apparent manifestation of this is the cosmic ray flux at ground level
which is dominated by muons created by the same pion decays that make atmospheric neutrinos.  The unstable pion is created by cosmic ray interactions in the diffuse upper atmosphere.  It then decays to make muons
and muon type neutrinos.  Many of the muons also decay adding to the
neutrino flux.  $\pi^{+} \rightarrow \mu^{+} + \nu_{\mu}$, $\mu^{+} \rightarrow e^{+} + \nu_{e} +\bar{\nu}_{\mu}$ and the charged conjugate reaction.

Intense energetic cosmic neutrino sources are expected to occur whenever energetic pions are produced in a region where pion decay is more likely than
absorption.  The observation of energetic cosmic gamma ray sources, powered
by neutral pion decay is strong evidence that point neutrino sources
should exist.

Neutrinos of several hundred MeV and up will interact with normal matter.
In particular the protons and neutrons in the nuclei of most matter provides
a convenient target.  When a charged current neutrino interaction occurs
at these energies there will be a charged lepton, an electron, muon or tauon
in the final state, with no apparent incident track since the neutrino is invisible until it interacts.  Neutrinos can also interact via the neutral
current interaction in which there is a neutrino in both the initial and final state.  Neutral current events have played an important role in
astrophysics, for specific reactions like deuterium dissociation, but in
general they do not provide enough information about the neutrino energy or the source direction.

As shown in Figure \ref{fig:zellertotalnumode} the neutrino cross section
increases approximately linearly at these energies.  The rise eventually
flattens out at the scale of the $W$ boson mass.  But for neutrino observations by this class of experiments it is a reasonable approximation.
\subsection{Motivation}
Progress in establishing neutrino astrophysics was a consequence of major
discoveries in high energy physics in the 1970's.  Gauge theories were
established as a cornerstone of the field with gauge theories for the strong,
weak and electromagnetic interactions.  A synthesis of these into grand
unified gauge theories predicted a quark lepton transition that would
make baryons, protons and neutrons, unstable.  Protons should decay.
The predicted lifetime was quite large, about $10^{28}$ years.  This was well
above established experimental limits on the proton lifetime.  To extend those
limits to the predicted lifetime would require massive detectors.
The technology for massive particle detectors had been well developed
during that time period.  Neutrino experiments were critical in establishing
the legitimacy of gauge theories and neutrino experiments require massive
detectors.  Such detectors would also be needed to establish grand unified
theories.

Neutrino detectors of that period where either small or very coarse, with
poor resolution.  Still some early experiments were based on existing
accelerator neutrino experiment technology.
\subsection{Detector Concept}
To observe proton decay one needed to see about one proton's mass of energy
deposited in a detector with very low momentum.  Since most massive detectors
are composed of nuclei the protons and neutrons are bound and so have some
Fermi momentum but it is well below the momentum one would have with a
proton's mass of energy.  For example a neutrino interaction depositing
one proton's mass of energy would deposit a comparable momentum.

\begin{figure}
	\centering
	\includegraphics[width=0.85\linewidth]{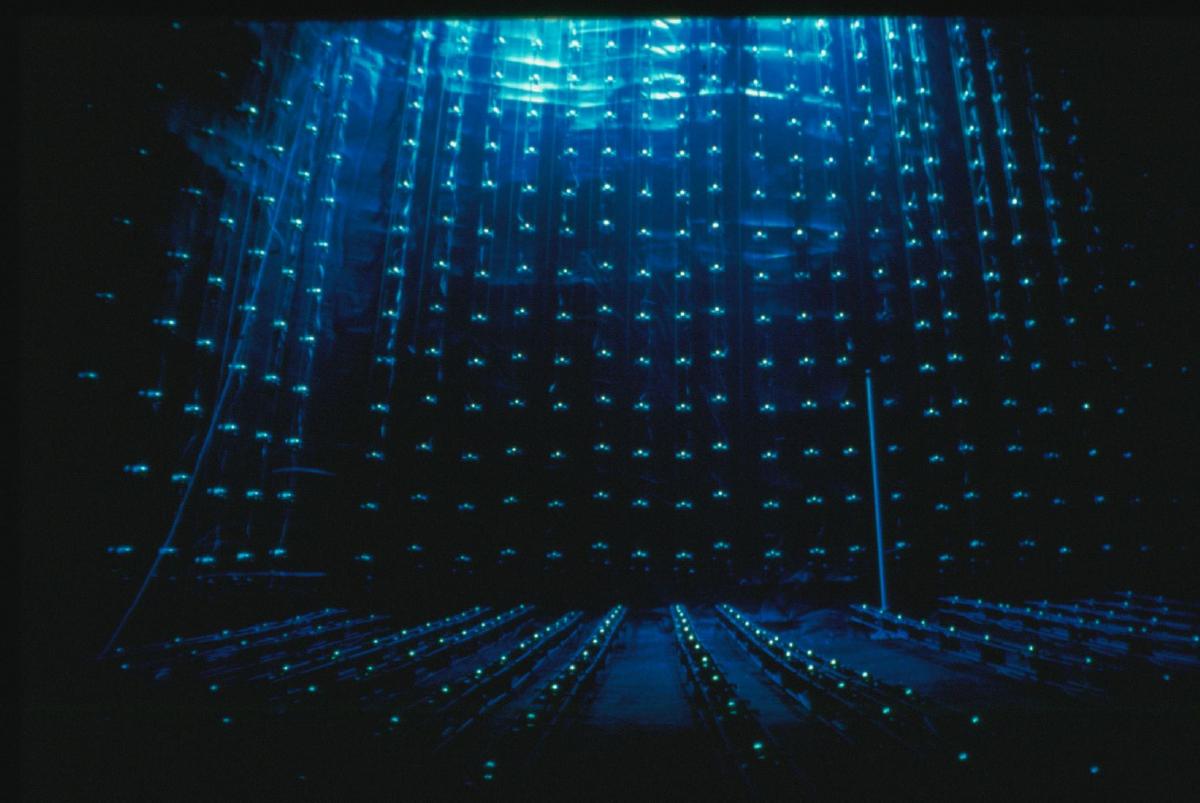}
	\caption{An underwater view of a wall of photomultiplier tubes in the original IMB detector.  The photo illustrates the large absorption length
		achieved for the water.  The distance scale is about 20 meters from wall to wall.}
	\label{fig:imbwall}
\end{figure}
To demonstrate a proton decay signal one had to present evidence for the
correct energy and momentum deposition.  Backgrounds that could be confused with proton decay were intrinsic radioactivity in the detector, cosmic rays
and neutrinos present in the environment.  Radioactivity can be managed
with careful construction avoiding certain materials.  Cosmic rays can
be managed by going deep underground.  Neutrinos can be recognized since
the momentum deposition would be comparable to the energy deposition.
A known source of background neutrinos comes from cosmic ray production
of unstable particles in the Earth's atmosphere.  These are called
atmospheric neutrinos and are the major background to many very sensitive
experiments on Earth.

A major breakthrough in detector design came in 1978 when the concept of
massive water Cherenkov detectors with surface light detectors would give
the energy and momentum information needed to distinguish proton decay
from backgrounds\cite{sulak}.  Surface light detection could be used if the water
was very clear with very little absorption or light scattering.  The
Cherenkov light rings projected onto the walls are used to reconstruct the
direction, length and the density of tracks in the interaction.
The diffuseness of the light can be used to classify the kind of track
as either showering (electron or photon) or non-showering (muon, pion etc.).

\begin{figure}
	\centering
	\includegraphics[width=0.495\linewidth]{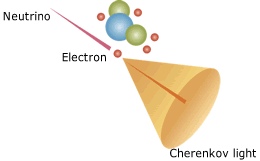}
	\includegraphics[width=0.496\linewidth]{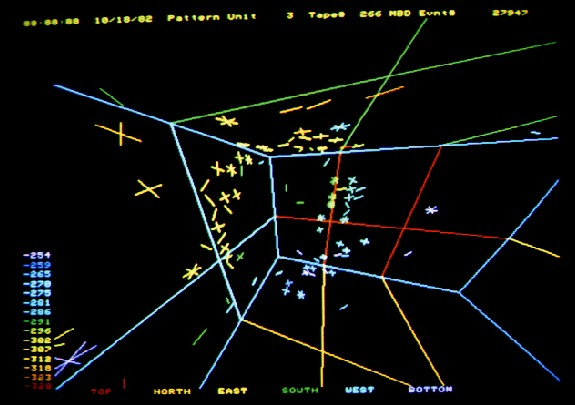}
	\caption{A sketch of the production of a Cherenkov cone (left). A computer display of the Cherenkov cone produced by a typical upward going neutrino interaction projected onto the walls of the IMB detector (right).}
	\label{fig:IMBUpNu}
\end{figure}

Cherenkov light is emitted by a charged particle moving through a transparent
material at a speed greater than the speed of light in that medium.  For water
the light is emitted in a cone at an angle of about 41.2$^{\circ}$ to the charged
track.  Figure \ref{fig:IMBUpNu} illustrates this and shows an example.

The detector concept can be traced back to DUMAND (Deep Underwater Muon And Neutrino Detector).
The DUMAND idea was to utilize
naturally occurring massive bodies of water acting as both neutrino target
and detector\cite{dumand1}.  The light detectors are distributed in the
volume of the detector and the optical quality of the water (or ice) is
hard to control.

An historic international meeting in Hawaii in 1976 the early
DUMAND concept was developed\cite{dumand2}. The Hawaii meeting had strong participation by many of the originators of the IMB detector (John Learned, Fred Reines and Lawrence Sulak).  The DUMAND concept has been used in a number of detectors to observe the
extremely low flux of very high energy neutrinos in cosmic rays.
This concept of using an imaging water Cherenkov detector has been used for many
astrophysical neutrino observatories including, IMB, Kamiokande, Super-Kamiokande and SNO.  More are planned or are under construction.
Ice Cube and related detectors are based on a similar concept, but use ice instead of water.

\section{Detectors}
\begin{figure}
	\centering
	\includegraphics[width=0.65\linewidth]{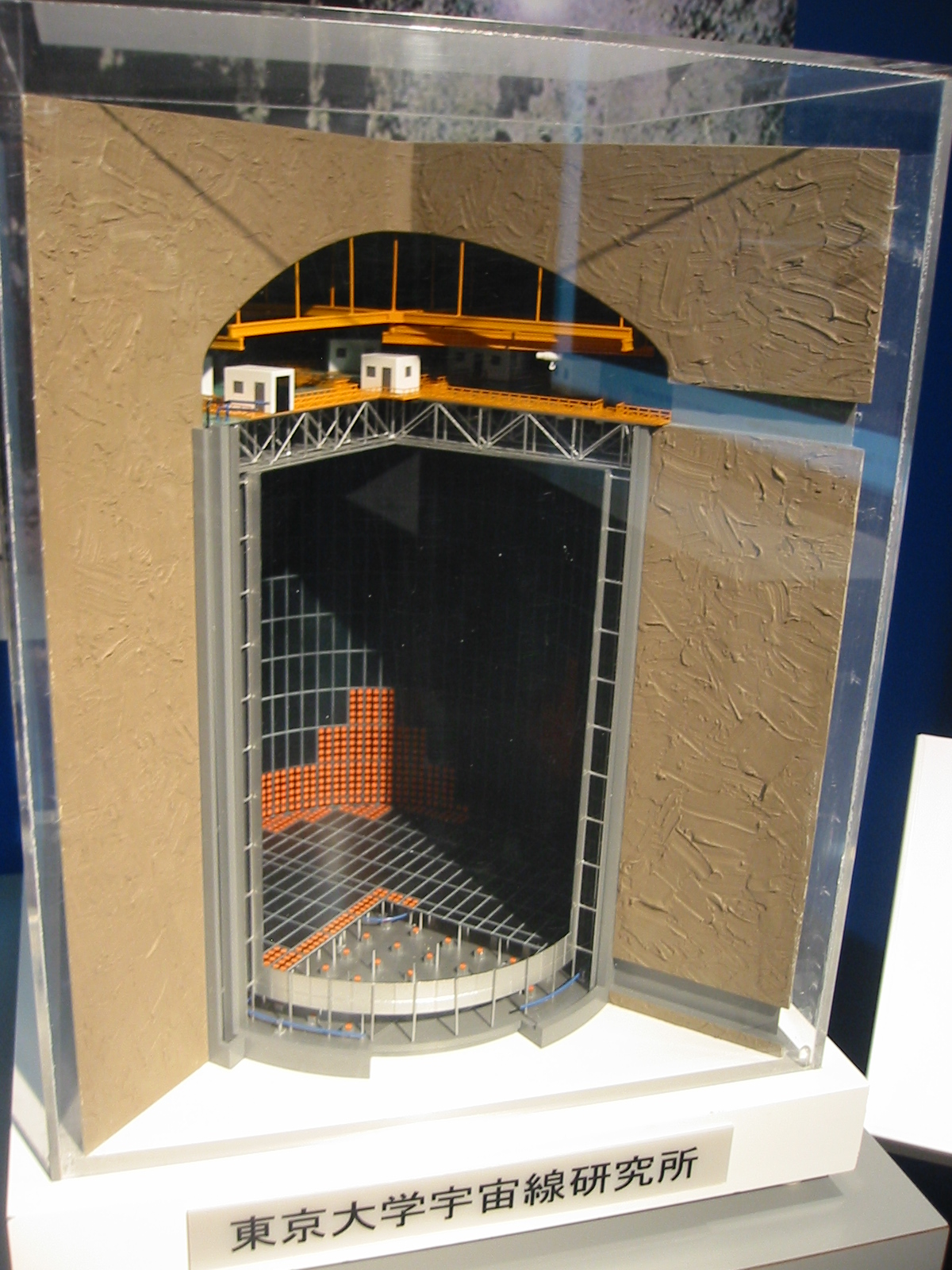}
        \includegraphics[width=0.65\linewidth]{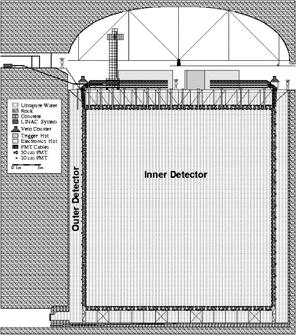}
	\caption{A model of the Kamiokande detector (top) and a drawing
          of the Super-Kamiokande detector (bottom).}
	\label{fig:Kam+SK}
\end{figure}
\subsection{IMB}
The first detector of this type was constructed by a collaboration between
the University of California at Irvine, the University of Michigan and
Brookhaven National Laboratory\cite{IMBDet}.  It was called IMB, the initials of the collaborating institutions.  Since the theory to be tested and the detector
technique were both speculative it was built ``on a shoestring''.  This was
literally true.  The light detectors were suspended in water with nylon
filament which were inexpensive and durable.  The use of the filament meant
that the detector could be maintained.  Failing components were lifted out,
repaired or replaced.  It also allowed the detector to go through two
upgrades to the light collection during its operation from 1979 to 1992.
The detector could adapt to new physics goals which resulted from its success
in achieving its initial goals.

The detector had a mass of 8000 tons of water and was viewed by 2048
photomultiplier tubes as light collectors.  Figure \ref{fig:imbwall}
shows the phototube layout and the clarity of the fluid with an under water
photograph.

Another feature of the first detector\cite{IMBDet}, IMB, was multi-hit electronics designed to
observe the decay of the proton decay products.  The proton decay products
would consist only of particles lighter than the proton.  Of these the kaon,
pion and muon are unstable and decay shortly after the proton.  Electrons and
photons are stable and so would not give a delayed time signature.

The initial proposal emphasized the goal of proton decay since it was topical
and had inspired the original design.  But the utility of such a device
in studying atmospheric neutrinos and supernova neutrinos was known and was mentioned in the proposal.  It was also known that the detector would be
sensitive to solar neutrinos.
\subsection{Kamiokande}
Kamiokande was a similar detector constructed in Japan in 1982.  The name comes from its location in the Kamioka mine in western Japan, plus Nucleon Decay Experiment (nde), signifying its original goal.  The detector
consisted of a cylindrical tank filled with 3000 tons of water viewed by
1000 20 inch photomultiplier tubes, figure \ref{fig:Kam+SK}.
\subsection{Super-Kamiokande}
Following the success of IMB and Kamiokande the two groups joined together
to construct a new generation detector.  The successes of the first round
devices helped focus the design on open physics questions.  The detector
is 50,000 tons of water in a cylindrical tank 40 meters high and 40 meters in diameter viewed by 13,000 20 inch photomultiplier tubes on the surface,
figure \ref{fig:Kam+SK}.  This review will not cover all the diverse results from this experiment, spanning solar neutrinos,
atmospheric neutrinos and neutrinos from a 300 km distant
accelerator source.  Many searches have been reported for
astrophysical phenomena but only the sun has yielded a positive
signal so far.
\subsection{MACRO}
The Monopole and Cosmic Ray Observatory (MACRO) in the Gran Sasso laboratory
of Italy was a large underground detector based on a segmented liquid
scintillation technology.  The detector was 76.6 m long, 12 m wide
and 9.3 m high subdivided into six sections called supermodules.
It was installed at a depth of about 3700 hg/cm$^2$.\cite{MACRODet}
The detector was composed of three subdetectors, liquid scintillation counters,
limited streamer tubes and nuclear track detectors.

The lower part of each supermodule had three horizontal planes and two vertical
planes of liquid scintillation counters.  The active volume of the horizontal
scintillators was 11.2m × 0.73m × 0.19 m viewed by two phototubes at each end.
The active volume of the vertical planes was 11.1m × 0.22 m × 0.46 m viewed by
one phototube at each end.  The total mass of the 476 scintillators was about
600 tons.

The lower part of the detector contained ten horizontal planes of limited
steamer tubes.  Between the middle eight layers was a rock absorber
with a thickness of about 360 g/cm$^{3}$ which set a threshold of about 1
GeV for a crossing muon.  Each tube was 12 M long and had a cross
sectional area of 3 cm x 3 cm.  The angular resolution of the streamer
tubes was about 0.2$^{\circ}$
The lower part of the detector contained one horizontal plane of the
nuclear track subdetector with additional planes on the east and north
faces.  There was a transition radiation detector in the upper
part of the device capable of measuring muon energies in the range of $100<E_{\mu}<930$ GeV.  Muon energies were also
estimated from multiple Coulomb scattering.

\begin{figure}[h]
        \centering
        \includegraphics[width=0.7\linewidth]{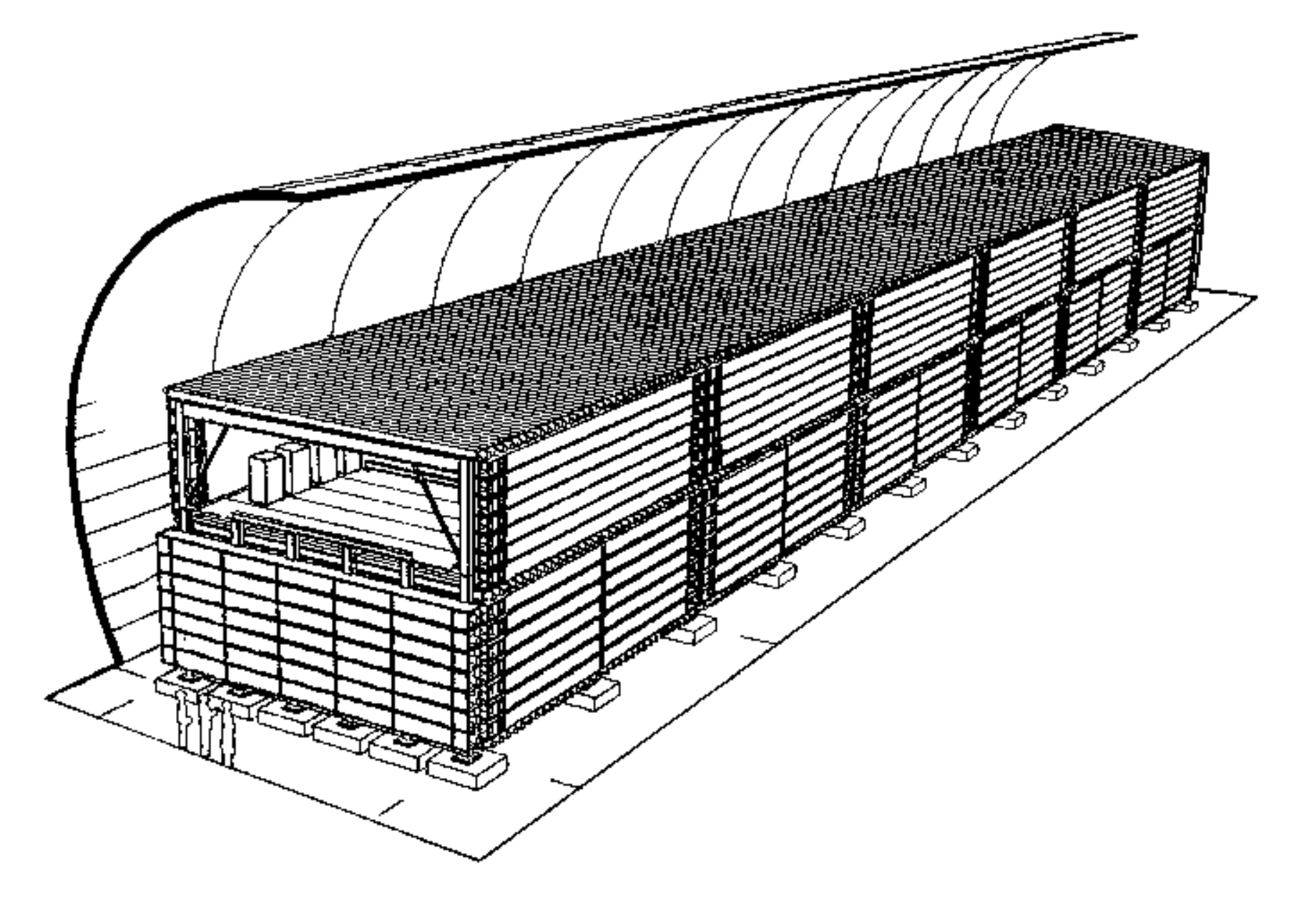}
        \caption{A sketch of the MACRO detector in the Gran Sasso Lab.
          Figure from \cite{GiacoandMarg}}
        \label{fig:MACROfig}
\end{figure}

Figure \ref{fig:MACROfig} is an illustration of the MACRO detector.
Unlike the imaging water Cherenkov detectors of the other devices
in this section MACRO was primarily a tracking device to detect and
identify signals from interactions outside the sensitive volume
of the detector.  Proton decay and direct neutrino observations were
not primary design goals.  Instead the detector had a very large sensitive
surface area and accurate timing to measure the speed and direction
of entering tracks.  Some of the physics goals included magnetic monopoles
in cosmic rays, atmospheric neutrinos and neutrino oscillations,
underground muons, high energy neutrino astronomy and indirect searches
for dark matter candidates.  They distinguish neutrino
interactions in the rock from penetrating cosmic ray muons by using time
of flight to distinguish the upward going muons, caused by neutrino
interactions, from the downward cosmic ray background.
\section{Neutrino Oscillations Complications}
Experimental neutrino physics primarily utilizes the charged current neutrino interaction as a detector and a source.  Neutrinos are usually created as
electron, muon or tau neutrinos as identified by the charged lepton involved
in the creation.  Most observations are via the charged current reaction
where the neutrino is identified via a charged lepton in the final state.
Neutral current neutrino interactions are observable but usually they lack the directional and energy information needed to understand their source.

Neutrino oscillations cause a propagating neutrino to change its flavor
from the one tagged at its creation.  This puzzled astrophysicists for decades
since reliable neutrino sources, such as the sun, could not be observed
as expected.  It took about 40 years to determine the cause of these
discrepancies as due to neutrino oscillations.

The detectors discussed in this review were key to resolving and understanding neutrino oscillations.  Starting with a muon deficit reported
by IMB\cite{NuAnom} to the observation of a distance dependent spectral distortion observed by Super-Kamiokande 15 years later\cite{Sk98}.

At astrophysical energies the problem was simple.  Electron type neutrinos
were created at the source as expected, but on route to the detector
some of them transformed into muon or tau neutrinos, which didn't have enough energy to engage in charged current interactions.  The possibility of sterile
neutrinos, neutrinos with no charged current interaction at all has also been
implicated by a number of experiments.

Detectors set up to observe neutrinos from the natural world lead the way
in understanding neutrino oscillations.  These conclusions have been confirmed
by terrestrial experiments with man made sources.

A full understanding of neutrino oscillations is needed to understand
neutrinos from astrophysical sources and must be accounted for.
From very large distances the neutrinos propagate as mass eigenstates
made up of superpositions of all the neutrino flavors.
\section{Results}
\subsection{Proton Decay}
To date there has been no convincing evidence for proton decay but the
search for proton decay continues to be a goal of most massive particle
detectors.  What had been a primary motivator has now become a secondary
goal.

The largest lower bound on  the proton lifetime come from Super-Kamiokande\cite{SKPDK}.  $\tau / Br(p \rightarrow e^{+} \pi^{0})>1.6 \times 10 ^{34}$ years with no candidate events observed and
$\tau / Br(p \rightarrow \mu^{+} \pi^{0})>7.7 \times 10 ^{33}$ years
with two candidates when 0.87 background were expected. 
\subsection{Atmospheric Neutrinos}
Atmospheric neutrinos, while a source of background to very sensitive experiments also presented an opportunity.  Up until that time the goal
of neutrino experiments was to maximize the signal.  To maximize the signal
one wanted the maximum neutrino flux.  So the detectors were placed as close to the source as possible.  The minimum distance was usually determined by radiation shielding issues since intense neutrino sources were also intense
radiation sources.
	\begin{figure}
	\centering
	\includegraphics[width=0.9\linewidth]{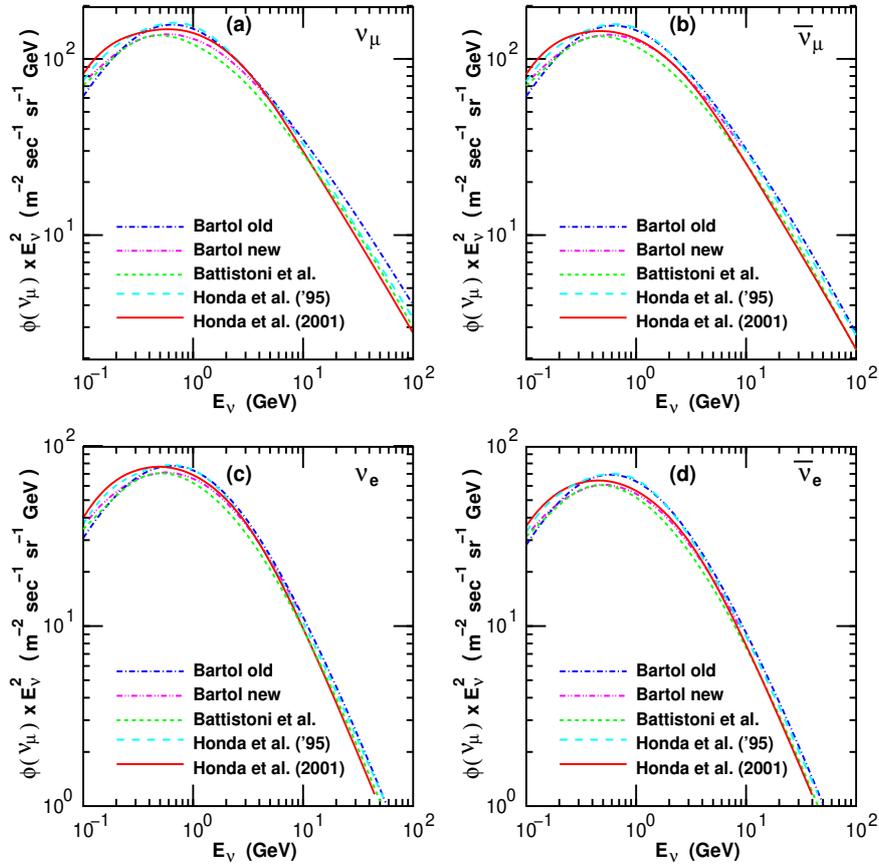}
	\caption{Estimates of the atmospheric flux (times $E^2$) for the various neutrino types. These estimates are made for the Kamioka location but are typical}
	\label{fig:gaisf18}
\end{figure}
	
The atmosphere surrounds the Earth and cosmic rays strike it from all directions.  Underground, or surface, detectors are closest to the
atmospheric neutrinos created near them but the flux is roughly isotropic.
The neutrinos come from all directions.
\begin{figure}
	\centering
	\includegraphics[width=0.7\linewidth]{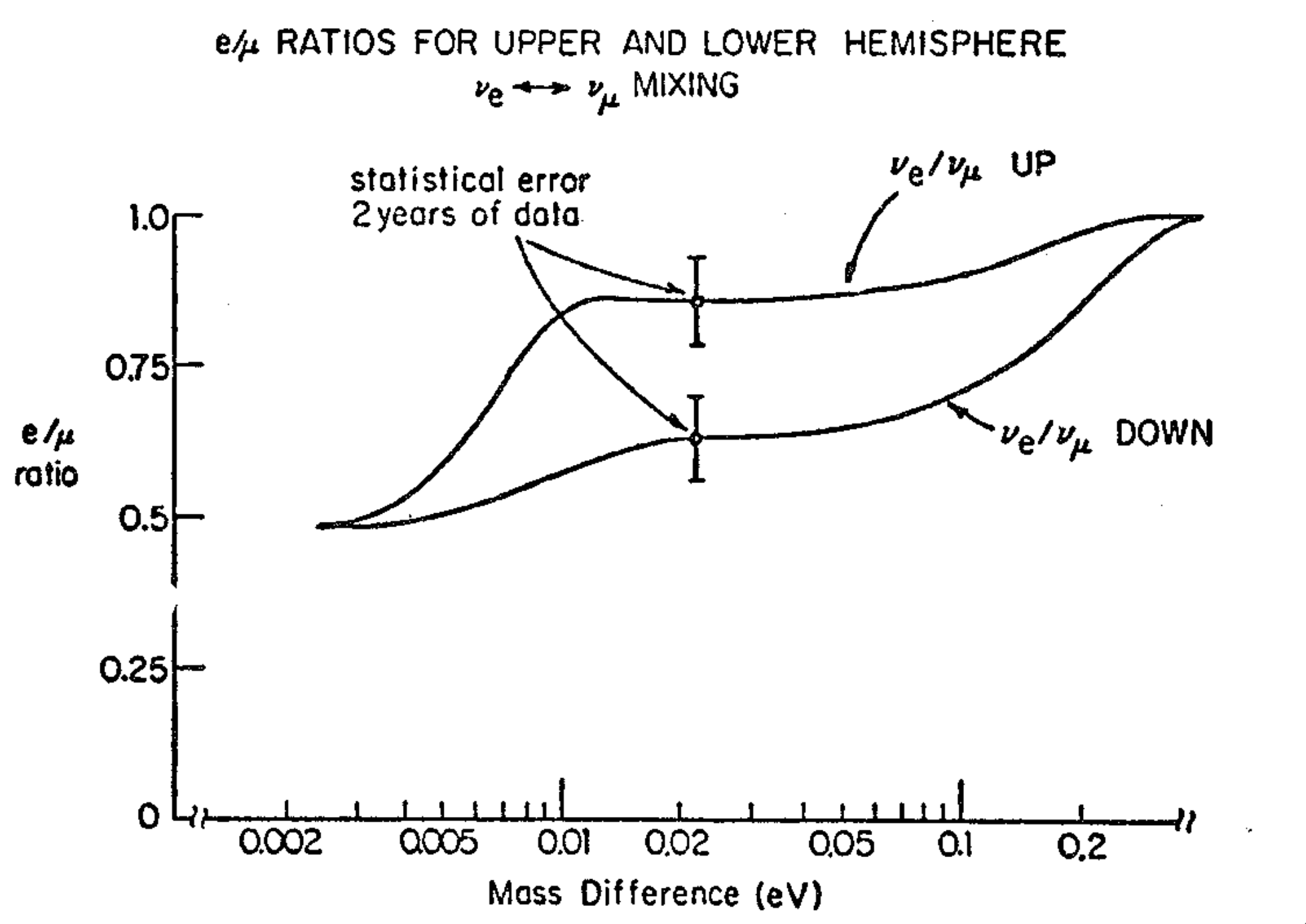}
	\caption{The ratio of electron neutrino interactions to muon neutrino
		interactions in the upward going and downward going hemispheres for
		maximal neutrino mixing as a function of the absolute neutrino mass
		difference, $\sqrt{\Delta m_{Atm}^{2}}$ The value rises from 0.5 for
		small values of $\Delta m_{Atm}^{2}$ to 1 for large values. The rise
		occurs at a much lower value of $\Delta m_{Atm}^{2}$ for upward neutrinos
		because those neutrinos travel two orders of magnitude further than
		downward ones.  Taken from \cite{Cortez}}
	\label{fig:fwogu-ne-nmu-3}
\end{figure}

Atmospheric neutrinos themselves provide little astrophysics information.
They are a tertiary result of the impact of cosmic rays on the earth.
The primary cosmic rays are charged.  Interstellar magnetic fields bend them
so they carry little information about their source.

The primary motivation to study atmospheric neutrinos is that they create
a background source for very sensitive experiments on earth that can not
be shielded.  The energy range for atmospheric neutrinos peaked in the energy
region near proton decay.  The motivation to understand the background to proton decay required an understanding of atmospheric neutrinos.
Figure \ref{fig:gaisf18} shows that the atmospheric neutrino flux tends to peak near the 1 GeV energy expected from proton decay.

The range of travel of atmospheric neutrinos varies from a few kilometers
for those from overhead to the Earth's diameter for those coming up from
below.  Without moving the detector one can study the time evolution of a neutrino beam over about three orders of magnitude.  One can compute the
travel time from the direction from which the neutrino has come.
\subsection{Neutrino Oscillations}
Neutrino oscillations are a property of neutrinos not specific to neutrino
astrophysics.  But understanding neutrino astrophysics was very limited
since oscillations of the neutrino signals while en-route made it impossible to understand the astrophysics at the source.

All of the detectors discussed in this section contributed to resolving this puzzle.  Until astronomers tried to observe neutrinos from distant sources
physicists had tried to study neutrinos by maximizing the flux by getting
very close to the source.  Neutrino oscillations are a quantum mechanical
time evolution of neutrinos that can not be noticed close to the source
since close to the source neutrinos do not have time to evolve.

In retrospect the first evidence for neutrino oscillations came from attempts
to observe the $^{8}B$ neutrinos from the sun.  Only about $\frac{1}{3}$ of
the expected flux was observed.  Neutrino oscillations were proposed as a solution but alternate explanations were also popular.  To a non-expert it
was also possible that the rare side reaction in the sun to make $^{8}B$
might have been overestimated or that the radiochemical method to extract
the signal argon from the chlorine might be inefficient.

Atmospheric neutrinos, an indirect consequence of cosmic rays in the atmosphere, provide a convenient source of neutrinos at a range of distances
from a few kilometers to the diameter of the earth.  One could compare a similar neutrino source at a range of distances from 10-20 km from those overhead to 13,000 km from those coming up.
Without moving the detector one can study the time evolution of a neutrino beam over about three orders of magnitude.  One can compute the
travel time from the direction from which the neutrino has come.
A simple two component model can describe the change in the flavor content
of an initially pure beam of muon neutrinos as the probability for the
$\nu_{\mu}$ to remain unchanged as:
\begin{equation}
P(\nu_{\mu} \rightarrow \nu_{\mu}) = 1-\sin^{2}(2 \theta_{Atm})\times \sin^{2}(1.27 \times \frac{\Delta m_{Atm}^{2} L}{E})
\end{equation}
where $\theta_{Atm}$ and $\Delta m_{Atm}^{2}$ are parameters and $L$ is a distance measured in km and $E$ is an energy measured in GeV.
$\Delta m_{Atm}^{2}$ is measured in units of eV$^{2}$.
\begin{figure}[th]
	\centering
	\includegraphics[width=0.85\linewidth]{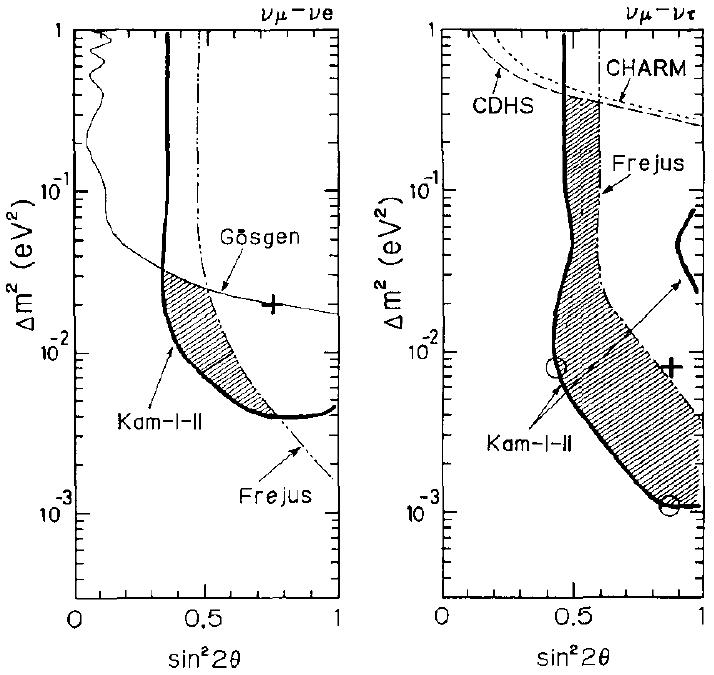}
	\caption{The Kamiokande 1992 fit to two component neutrino oscillations.
	The figure on the left considered $\nu_{\mu}\rightarrow\nu_{e}$ oscillations which were subsequently ruled out by a reactor $\bar{\nu}_{e}$ experiment}
	\label{fig:hirataatmnufitpl1992oscfit}
\end{figure}
The atmospheric neutrino energy spectrum has modest directional energy dependence, due to the earth's magnetic field, that can be calculated.  The neutrino oscillations parameters can be estimated from the variation of the muon neutrino flux with direction, $L$ to find $\Delta m_{Atm}^{2}$ and the magnitude of the modulation to find $\theta_{Atm}$.

Proton decay experiments studied atmospheric neutrinos to understand the background to proton decay but also realized their potential to be used
to study neutrino oscillations\cite{Cortez}.  This source of neutrinos is created as a mixture of four kinds of neutrinos, $\nu_{\nu}, \nu_{e}, \bar{\nu}_{\nu}, \bar{\nu}_{e}$, figure \ref{fig:gaisf18}.

Given the range of neutrino propagation distances and the neutrino energies
only a range of  $\Delta m_{Atm}^{2}$, from about $10^{-4}$ to $10^{-2}$ $eV^{2}$, will provide substantial modulation of the neutrino flux.  for values much smaller than
this the neutrinos will not have time to evolve so the observations will match expectations.  For values much greater than this the neutrinos will be fully mixed and will show little
modulation.  In this second case the neutrino type ratios measured would
deviate from those expected from the flux estimates.

A simple argument suggests that atmospheric neutrinos should have twice as
many muon neutrinos as electron neutrinos.  The neutrinos originate from
pion decay to a muon and a muon neutrino.  The muon is unstable and will decay to a muon neutrino, an electron neutrino and an electron.
$\pi^{+}\rightarrow\mu^{+}+\nu_{\mu}$ and $\mu^{+}\rightarrow e^{+}+\bar{\nu}_{\mu}+\nu_{e}$.
So one gets two muon type neutrinos for every electron neutrino.
This argument is a bit sloppy since the neutrinos produced will have different energies and so different interaction rates.  The neutrino and anti neutrino interaction rate on ordinary matter differ.

\begin{figure}
	\centering
	\includegraphics[width=0.6\linewidth]{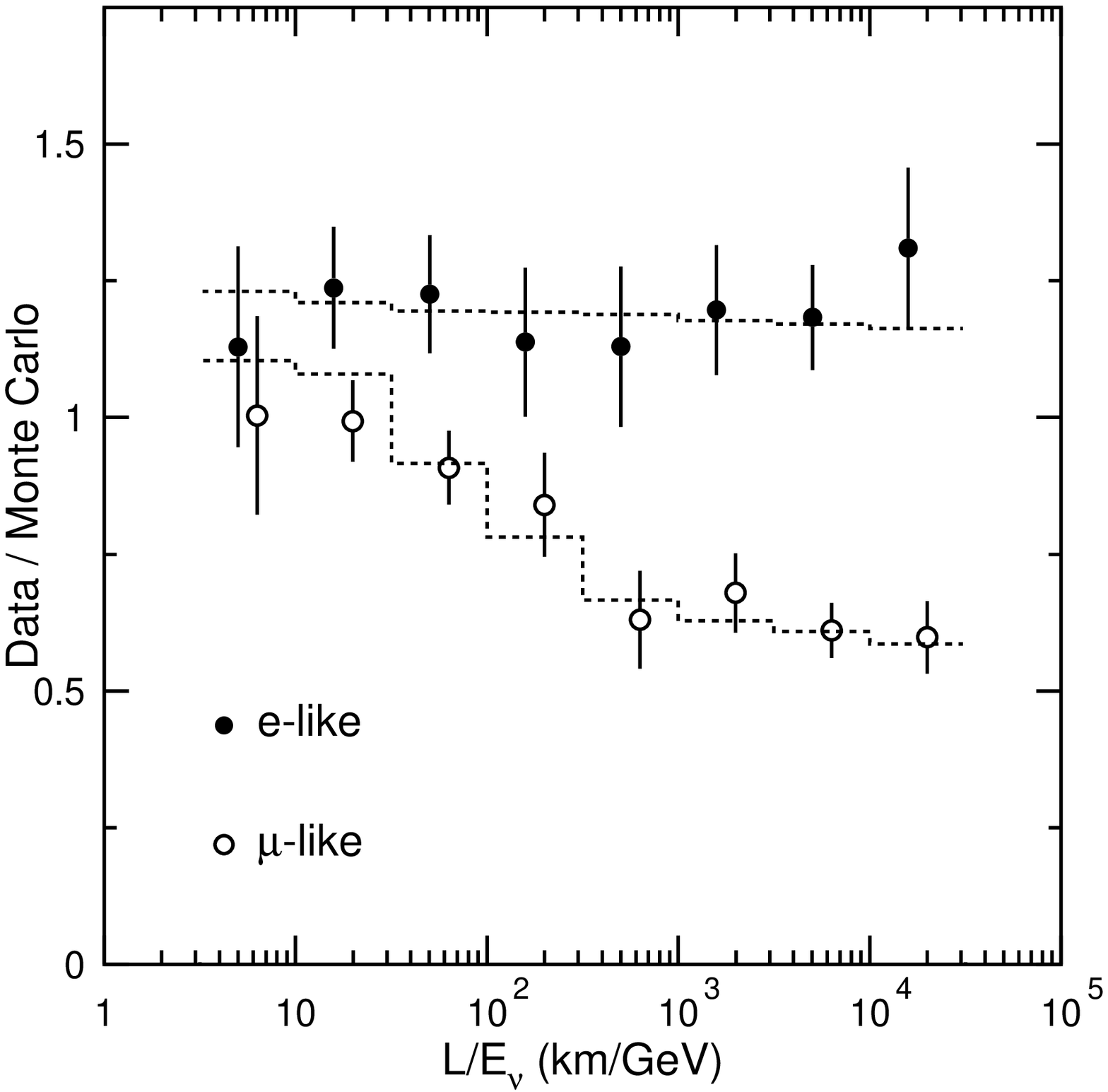}
	\includegraphics[width=0.6\linewidth]{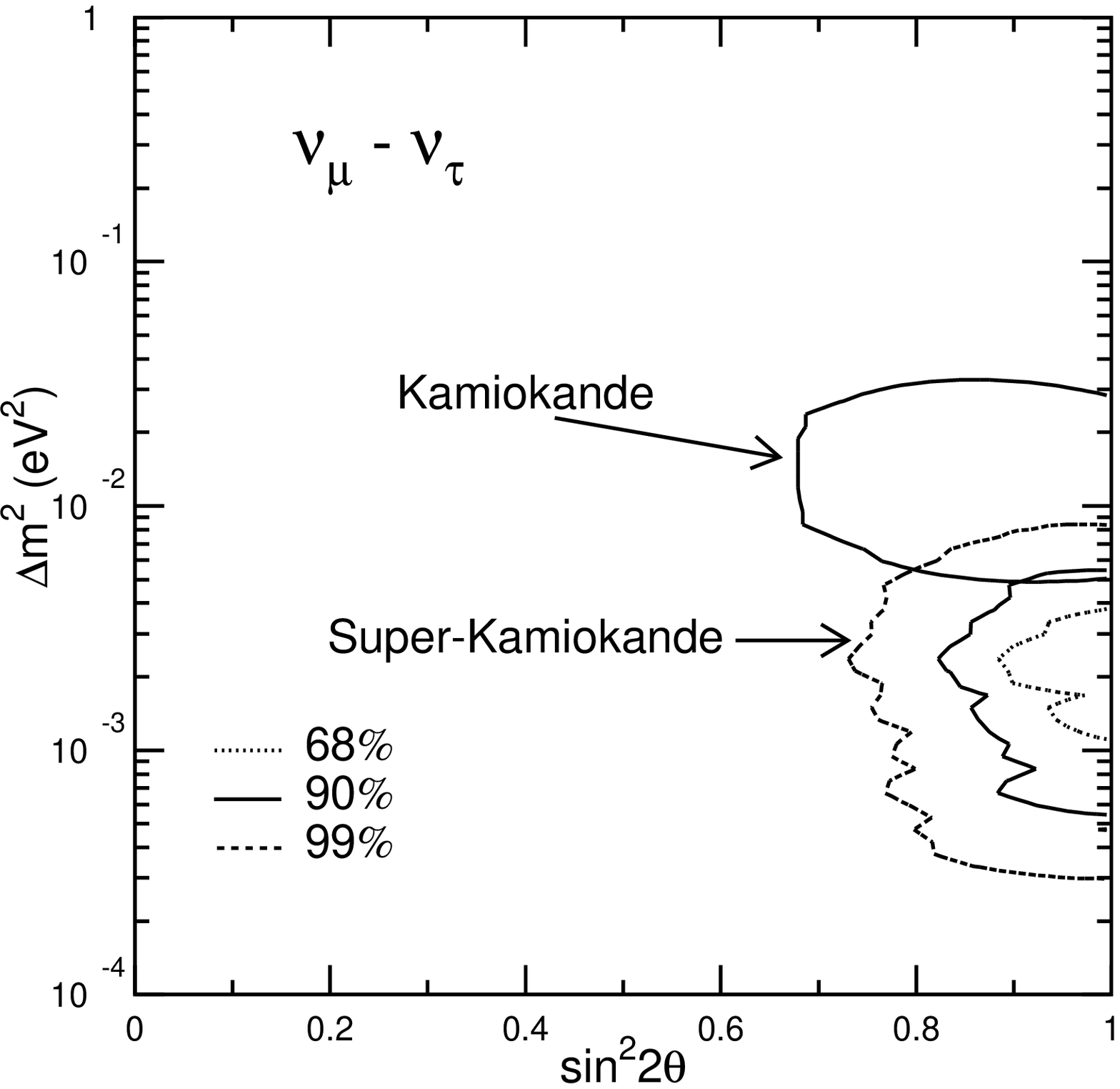}
	\caption{The Super-K 1998 L/E distribution for electron and muon events (top).
The Super-K 1998 fit of the atmospheric neutrino data to oscillations (bottom).
The fit involves 8 parameters in addition to the two for neutrino oscillations}
	\label{fig:sk98kamsk}
\end{figure}
The first evidence that the observations did not match the expected flux came from IMB where the earliest results showed a deficiency of muon neutrino
interactions as a fraction of the total measured sample\cite{BruceBill}.
These initial indications were followed up with a thorough search for possible
causes for the missing muons.  Experimental problems were ruled out since
the easily identified sample of muons from cosmic rays that stopped in the detector behaved as expected.  An IMB comparison of an upward going sample with a downward going sample set limits on neutrino oscillations\cite{NuOscPRL1985}.  A neutrino oscillation analysis using only
muon neutrinos and a plot of the neutrino $E/L$ distribution were also
published\cite{ICRCIMB1985}.
No claims of discovery were made at the time
because other experiments using different muon end electron identification
methods were indicating no anomaly at the time\cite{NuSex,Koshiba}.

In early 1986 the IMB muon neutrino deficit was compared with results
of NuSex and Kamiokande in a review talk by LoSecco \cite{LLou}
``Both the Kamioka detector and the Nusex detector can distinguish
$\nu_{e}$ from $\nu_{\mu}$ by shower development.  They quote a $\nu_{e}/\nu_{\mu}$ flux ratio of 0.36$\pm$0.08 and 0.28$\pm$0.11 respectively.  These are lower than the expected value of 0.64.  The IMB group has studied the fraction of their contained events resulting in a muon decay. The 26\% observed can be converted to a $\nu_{e}/\nu_{\mu}$ ratio with
a number of assumptions about muon capture in water.  If 40\% of the $\nu_{\mu}$ interactions do {\em not} result in a muon decay signal the observed value corresponds to $\nu_{e}/\nu_{\mu}\approx1.3$.  The problem of the $\nu_{e}/\nu_{\mu}$ ratio is still under active study.''.  The three experiments used different methods to make their measurements.  LoSecco had
noticed earlier, from a table published by Kamiokande\cite{Koshiba}, that while their official result based on shower development showed the quoted electron deficiency their muon decay rate supported the IMB result.
Shortly after the review talk the IMB deficit was published in Physical Review Letters\cite{NuAnom}.  In June of 1986 LoSecco traveled to Tokyo to discuss the discrepancy between the two classifications used by
Kamiokande.  He was told unambiguously that the shower development method,
yielding a muon excess, was correct.   The full story of the eventual confirmation of the IMB deficit by Kamiokande
in 1988\cite{Hirata} can be read here \cite{AnomHist}.
\begin{figure}
	\centering
	\includegraphics[width=0.75\linewidth]{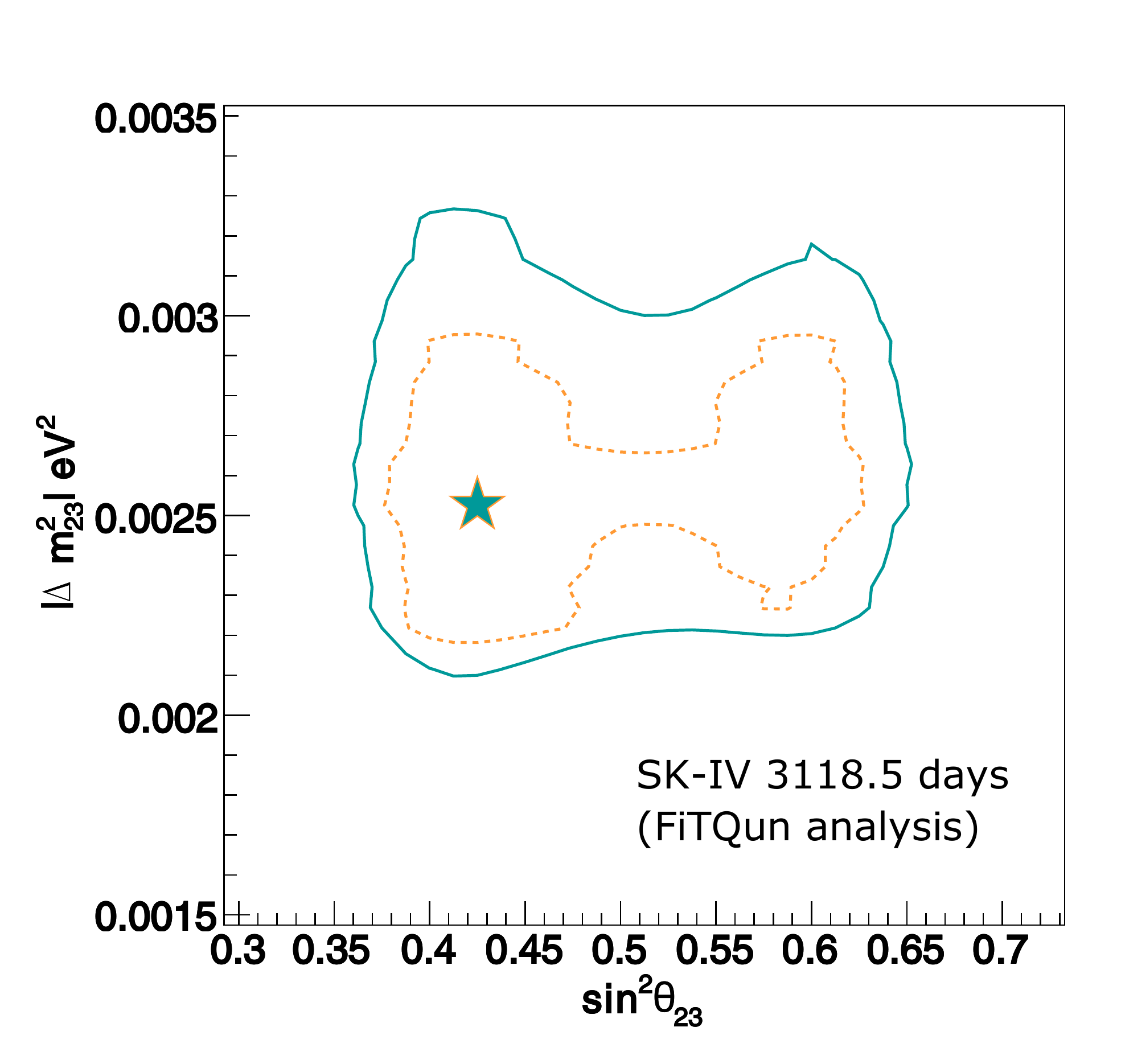}
	\caption{The Super-Kamiokande IV fit to $\Delta m^{2}_{23}$ and $\sin^{2}{\theta_{23}}$}
	\label{fig:sk417astandardq13-fixedfullsinglenhihs23m23}
\end{figure}

The effect was large.  About 40\% of the expected muon signal was missing.
Attempts to fit the neutrino oscillation hypothesis was stymied by the absence
of any significant variation with distance or energy.  The best fit to
these data, figure \ref{fig:hirataatmnufitpl1992oscfit}, gave $\Delta m_{Atm}^{2} = 8 \times 10^{-3}$ eV$^{2}$ and
$\sin^{2}(2 \theta_{Atm})=0.43$, basically isotropic with 40\% of the muons missing.  Since the oscillation probability depends
on $\frac{L}{E}$ the absence of dependence on either of these was problematic.
The distance traveled $L$ is limited by the size of the earth.  On the other hand the energy $E$ was limited by the detector size.  If an neutrino interaction entered or exited detector one only had partial information about
its energy.  Entering tracks had an additional advantage since these tracks,
if going upward, came from neutrino interactions in the rock.  The rock represented a much larger mass neutrino target so the neutrino interaction rate would be larger.  Still a lot of information was missing.  Attempts
to understand the muon deficit using upward going stopped and through going
events were limited by modeling of the neutrino source and the interaction volume.  Ultimately a larger detector was needed to resolve the question.
\begin{figure}[tb]
	\centering
	\includegraphics[width=0.9\linewidth]{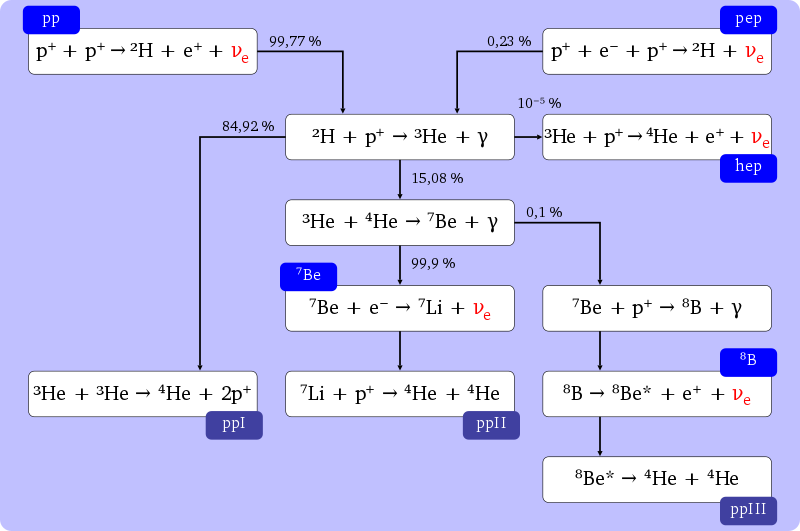}
	\caption{Nuclear reactions yielding neutrinos in the sun.  The neutrinos
		are labeled in red.  Branching ratios are shown}
	\label{fig:snreactions}
\end{figure}
\begin{figure}[tb]
	\centering
	\includegraphics[width=0.9\linewidth]{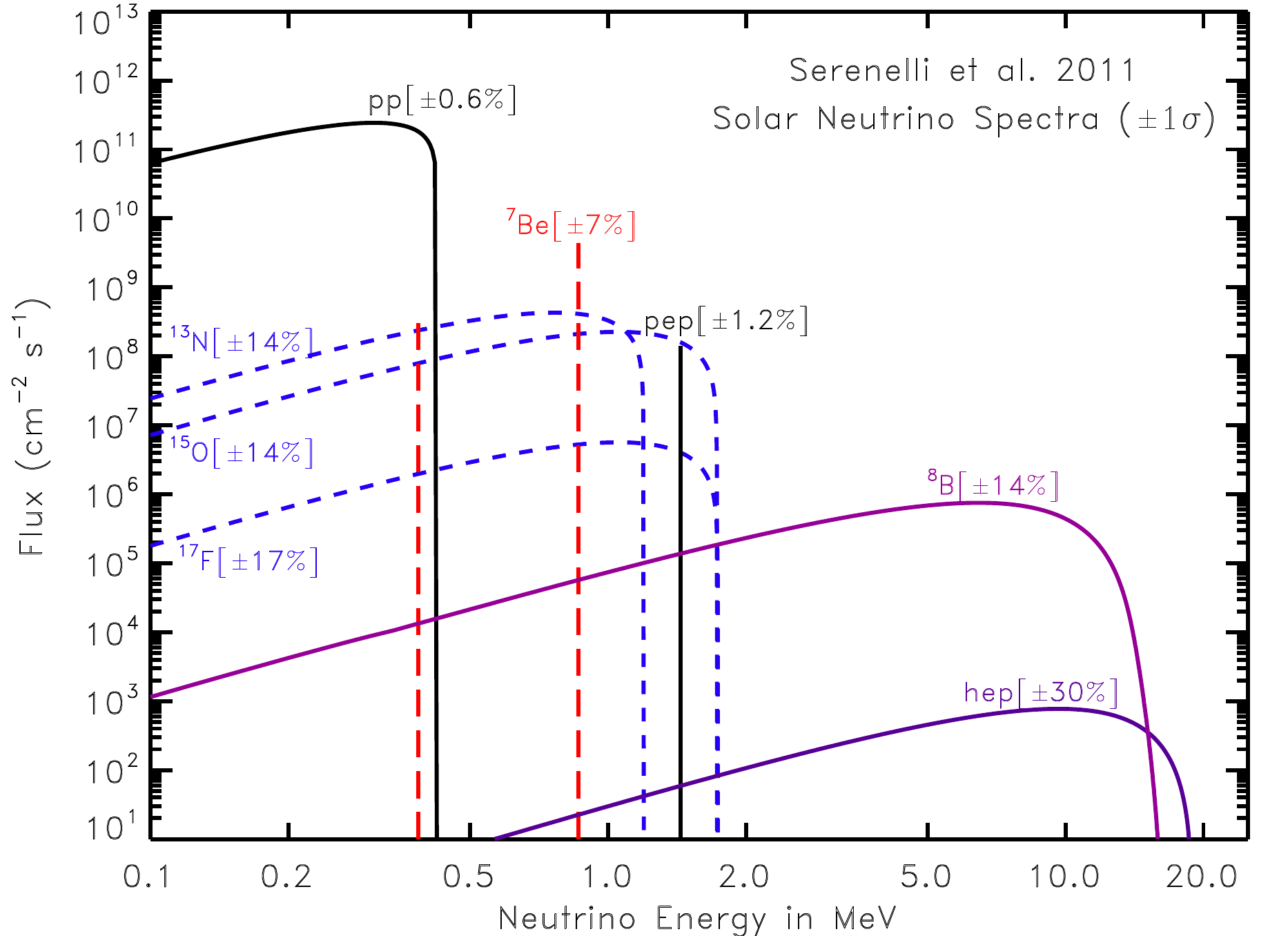}
	\caption{The solar neutrino fluxes.  Taken from \cite{snu}}
	\label{fig:snfig3fluxes}
\end{figure}
\begin{figure}[tb]
	\centering
	\includegraphics[width=0.7\linewidth]{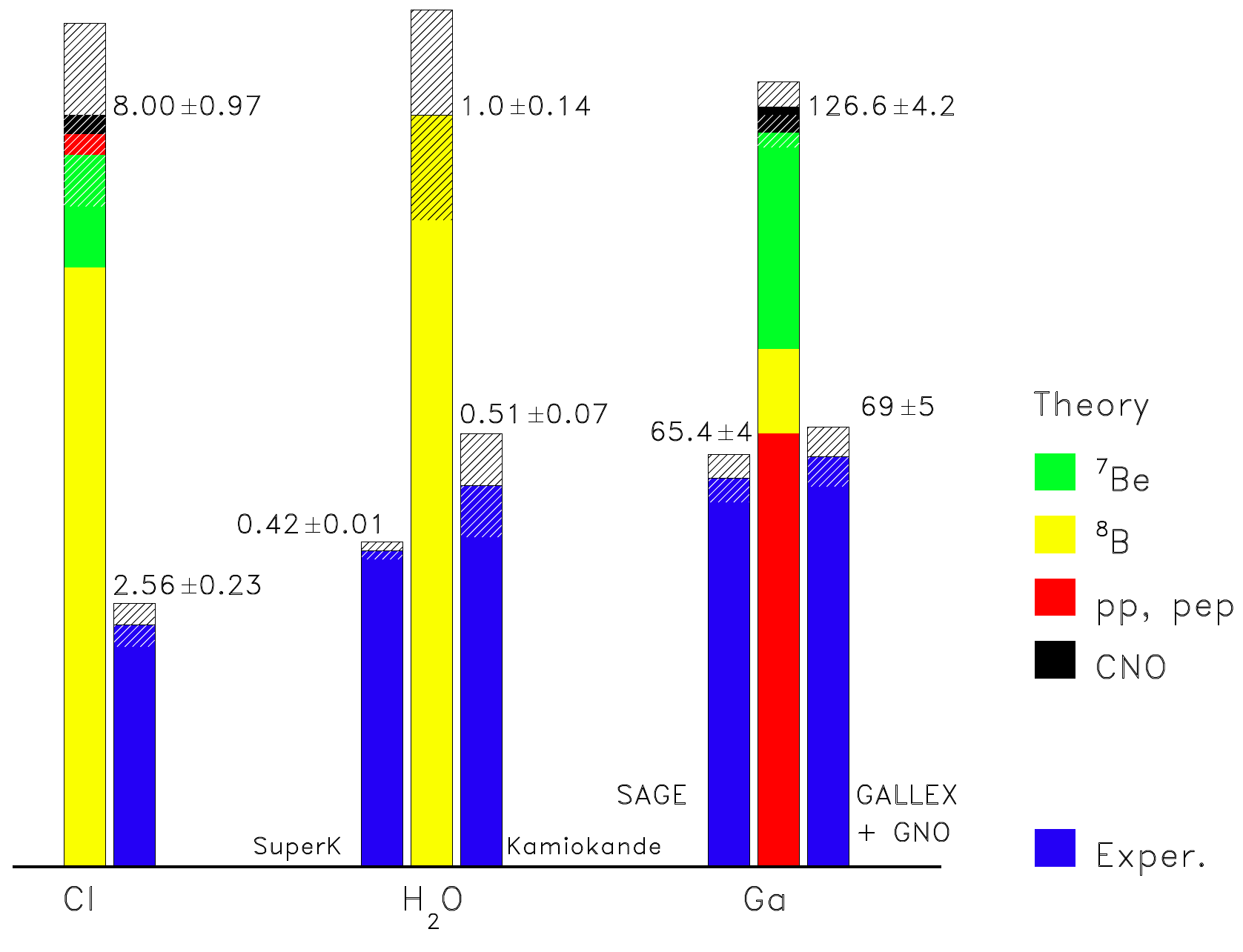}
	\caption{The solar neutrino observations.  Taken from \cite{snu}}
	\label{fig:snfig3fluxesObs}
\end{figure}

A strong motivation for the 25 kiloton Super-Kamiokande detector was to extend
the energy range over which neutrino energies could be measured.  Super-K
began observations in 1996 and by 1998 had accumulated enough atmospheric
neutrino interactions, almost 5000 events, to make a clear measurement\cite{SuperK98}.
\begin{figure}[tb]
	\centering
	\includegraphics[width=0.875\linewidth]{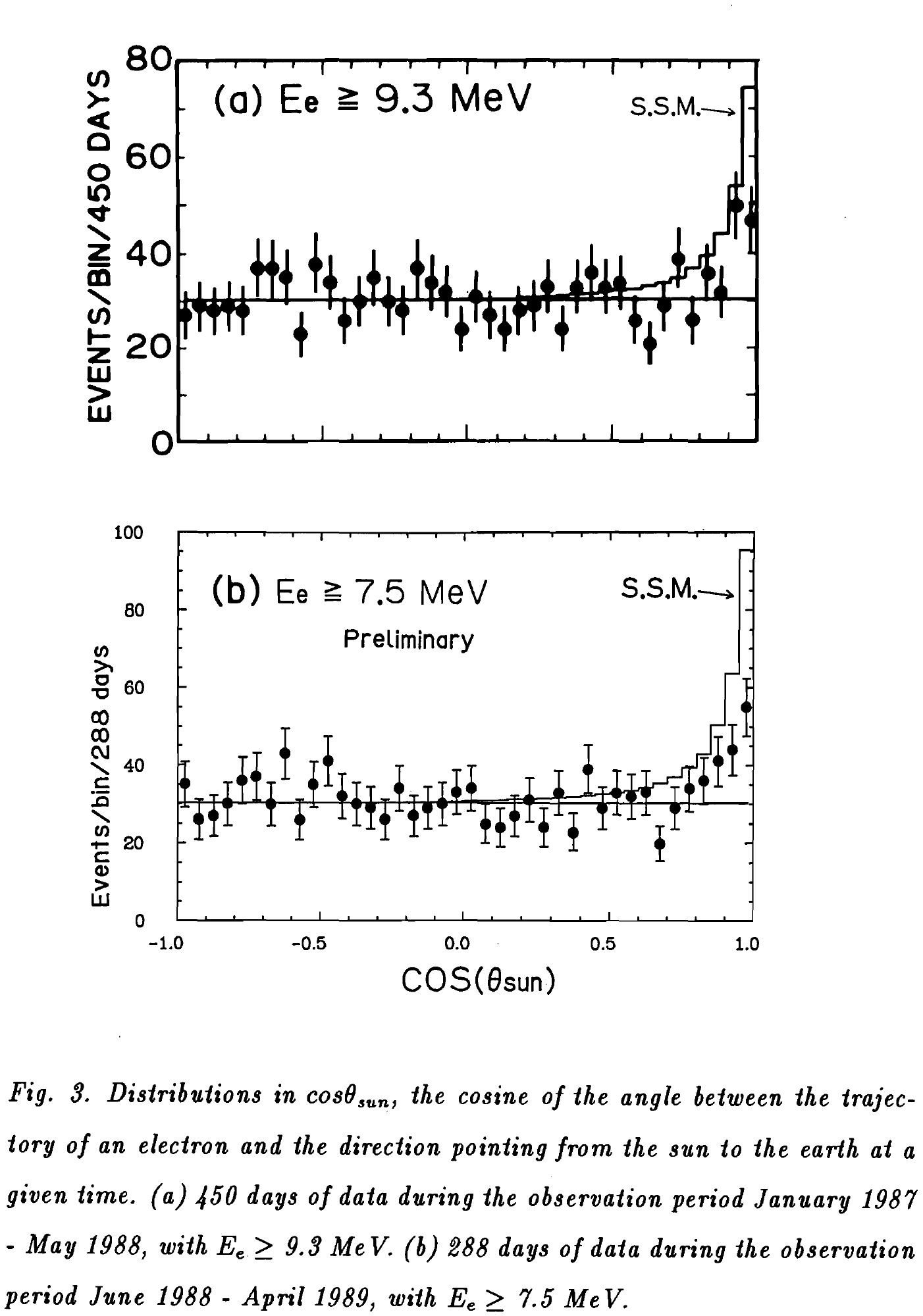}
	\caption{The Kamiokande results on solar neutrinos, 
		The angular distribution.
	 Taken from Nakamura\cite{KamStat}.}
	\label{fig:kamikandesolnu}
\end{figure}
\begin{figure}[tb]
	\centering
	\includegraphics[width=0.7\linewidth]{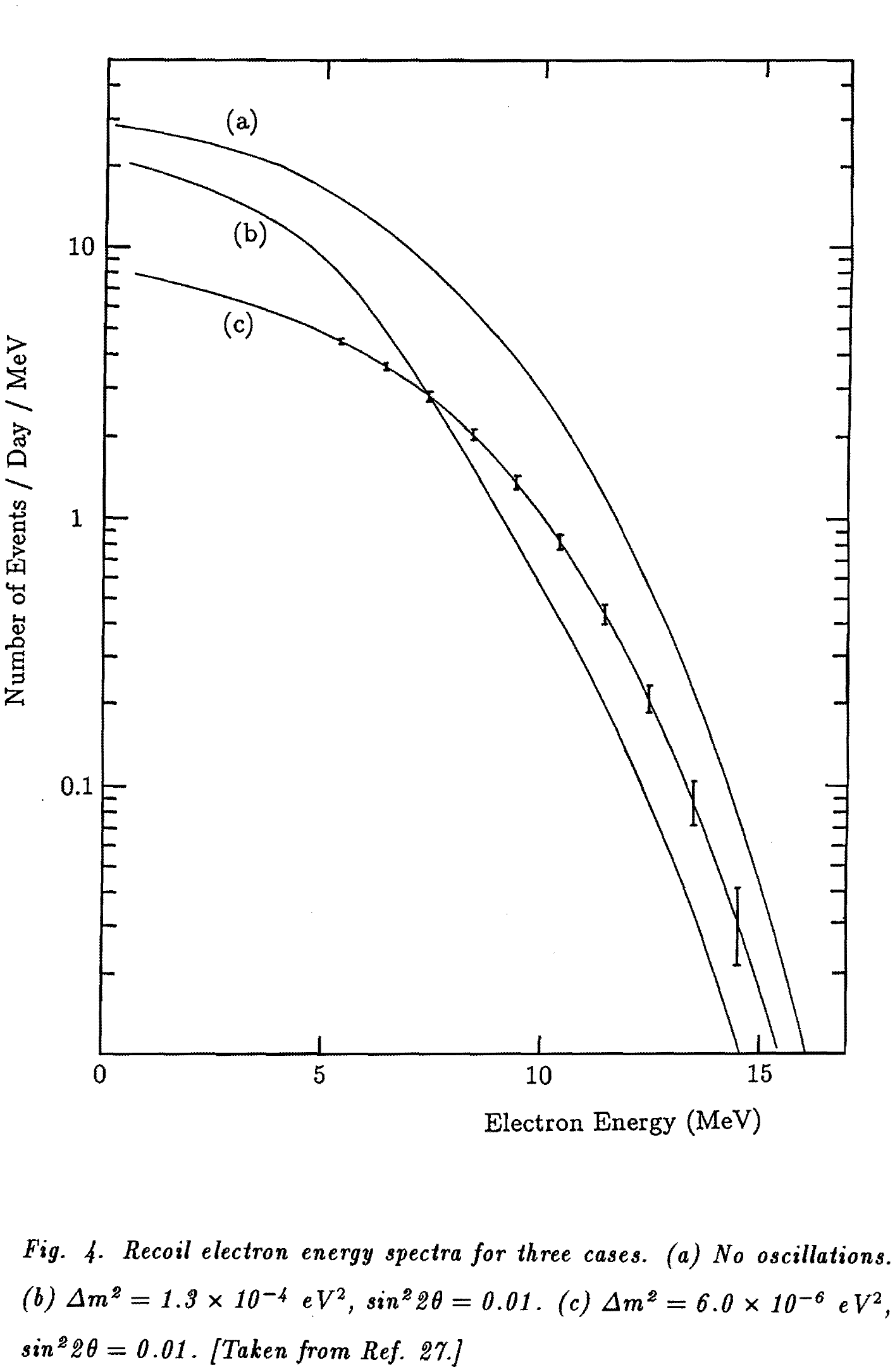}
	\caption{The Kamiokande results on solar neutrinos. The energy distribution is compared to the standard solar model and two models of neutrino oscillations.  Taken from Nakamura\cite{KamStat}.}
	\label{fig:kamikandesolnuE}
\end{figure}

The most recent Super-Kamiokande report on atmospheric neutrino
oscillations\cite{SuperK19} was based on a 253.9 kton-year exposure and had
about 24,000 events.  The fit over 515 analysis bins, figure \ref{fig:sk417astandardq13-fixedfullsinglenhihs23m23}, results in $\Delta m_{32}^{2} = 2.63 \pm ^{+0.10}_{-0.21} \times 10^{-3}$ eV$^{2}$ and $\sin^{2}{\theta_{23}}=0.588^{+0.030}_{-0.062}$ or $\sin^{2}{\theta_{23}}=0.425^{+0.051}_{-0.034}$ in the first octant with normal mass hierarchy.  The fit includes 118 free parameters in addition to those describing neutrino oscillations.

MACRO was able to measure atmospheric $\nu_{\mu}$ interactions
via three channels: upward through going muons due
to $\nu_{\mu}$ interactions in the rock with a mean neutrino
energy of
50 GeV, upward going exiting muons with a neutrino interaction
in the lower part of the detector and upward stopping muons
due to a lower energy $\nu_{\mu}$ interaction in the rock.
The average neutrino energy for the last two channels was
2-3 GeV.  The upward through going muons had an angular distribution inconsistent with that expected from no neutrino
oscillations.  The angular distribution is explained by
$\nu_{\mu}\rightarrow\nu_{\tau}$ with parameters consistent
with Super-Kamiokande quoted above.\cite{MACRONuOsc}  Due to substantial uncertainties from the neutrino flux estimates subsequent analyses
using 5 ratios of signal sub-samples were done.  The ratio of
vertical to horizontal high energy muons, the ratio of
low energy to high energy through going muons, the ratio
of the ratios of data to Monte Carlo for the upward going
neutrino events interacting in the detector to the ratio
of interacting downward going and upward going stopped events, the ratio of the observed over expected through going muons
and the ratio of low energy semi-contained events (starting
or ending in the detector) over the expected value.
All of these ratios confirmed the shape fit.
\subsection{Solar Neutrinos}
The sun and most stars derive their energy by converting hydrogen into
helium.  For this to happen two protons must convert into neutrons, creating
a positron and an electron neutrino for each conversion.  The neutrinos escape
the core of the sun.  There are additional nuclear reactions in the
sun and stars that extend the neutrino energy well above what one would
have with simple hydrogen "burning".

The proton-proton cycle converts hydrogen into helium in the suns core\cite{snu}.
\[ 2 e^{-} + 4 p \rightarrow ~^{4}He + 2 \nu_{e} + 26.73 MeV\]
It produces two $\nu_{e}$ and 26.73 MeV of energy for every He produced.
The energy produced thermalizes and the thermal pressure balances the
gravitational pressure to provide hydrostatic equilibrium.  The energy produced at the center of the sun eventually is radiated from the surface
and must be replenished to keep the sun in thermodynamic equilibrium.
The neutrinos produced leave the core and can be observed.

The full proton proton chain has several steps with intermediate products
\[p + p \rightarrow ~^{2}H + e^{+} + \nu_{e}\]
creates $~^{2}H$, deuterium.
\[p + ~^{2}H \rightarrow ~^{3}He\]
creates $~^{3}He$, helium 3 and
\[~^{3}He + ~^{3}He \rightarrow ~^{4}He + 2 p\]
releases the two intermediate protons.

The sun is consuming fuel and so its energy generation is evolving.  The time
scale for changes is billions of years, so for observational work it may
be considered a static source.

The neutrino flux from the proton proton reaction, $ p + p \rightarrow ~^{2}H + e^{+} + \nu_{e}$, is about $6.05 \times 10^{10} \nu/cm^{2}/s$ at earth.
But the endpoint of this reaction is about
420 keV which makes it hard to observe except with the radiochemical
reaction $\nu_{e} + ~^{71}Ga \rightarrow ~^{71}Ge + e^{-}$.  The spectral
shape is known so an integrated reaction rate can be computed from the flux
and cross section above the $~^{71}Ga$ reaction threshold of 233 keV.

The reaction $~^{3}He + ~^{4}He \rightarrow ~^{7}Be + \gamma$ opens up
several opportunities for nuclear reactions yielding higher energy
neutrinos.  In particular the reaction $~^{7}Be + p \rightarrow ~^{8}B + \gamma$ provides access to the $~^{8}B$ decay neutrinos
$~^{8}B \rightarrow ~^{8}Be + e^{+} + \nu_{e}$
which extend out to
energies of about 15 MeV as seen in figure \ref{fig:snfig3fluxes}.
The process $~^{7}Be + e^{-} \rightarrow ~^{7}Li + \nu_{e}$ produces
two energies at 860 keV (90\%) and 380 keV (10\%).  The highest energy
neutrinos come from the reaction $~^{3}He + p \rightarrow ~^{4}He + e^{+} + \nu_{e}$ with an endpoint energy of 18.7 MeV.  But the reaction is very rare
and yields a flux of about $8 \times 10^{3}\nu/cm^{2}/s$ at earth.

Many techniques of solar neutrino observations use radiochemical methods
such as $\nu_{e} + ~^{37}Cl \rightarrow ~^{37}Ar + e^{-}$ or
$\nu_{e} + ~^{71}Ga \rightarrow ~^{71}Ge + e^{-}$ where the unstable
final state nucleus is extracted and observed via its own decay.
Such methods can measure an integrated flux but give no information about
the neutrino energy or direction.  The $~^{71}Ga$ reaction has a threshold of
233 keV, below the 420 keV endpoint of the proton proton reaction.
The produced $~^{71}Ge$ has a half life 
of 11.43 days and can be observed via its decay.  Because $~^{71}Ga$ observes the
initial reaction of the main energy production mechanism in the sun
one expects about 2.69 reactions with solar neutrinos in 10 tons of
natural $Ga$ during the 11.43 day lifetime of the $~^{71}Ge$
Because of their low energy thresholds radiochemical methods have played a
critical role in understanding solar neutrinos.  But the methods only measure an integral and are insensitive to neutrino flavor.  The very low rate of
solar neutrino events in $~^{37}Cl$ (figure \ref{fig:snfig3fluxesObs}) was the initial solar neutrino problem
which took decades of work to resolve.

Water contains oxygen  nuclei, free protons and electrons.  Electron neutrinos
from the sun will not have charged current reactions on the nuclear targets present.  Other neutrino flavors also have insufficient energy to react
with these nuclei.
So the water detectors covered by this review utilize neutrino electron scattering, which provides directional information.

IMB considered studying solar neutrino scattering.  It was suggested by
Tegid W. Jones of University College London when he joined the collaboration and moved temporarily to Michigan.  Jones had been a member of the Garagamelle
collaboration that had first observed $\nu + e^{-}$ scattering.
IMB concluded that at its depth
of 600 m, about 1570 meter water equivalent the decay of muon produced
radionuclides would make it hard to see a solar signal.  The muon rate at IMB was 2.73 muons per second.  Observing solar neutrino scattering would have
required much more light collection to get down to the 5-10 MeV energy
region where the signal would be found.  At those energies radioactive contaminants are also significant and would have required significant remedies
to eliminate.  On the plus side, salt mines are much cleaner radiologically
and the mine had a tradition of doing very low background nuclear physics,
such as double beta decay.
Administratively, proton decay was funded by the high energy community but
solar neutrinos was identified with the nuclear physics community.  It wasn't clear if the nuclear physics community would support US work, in this manner at that time.
\begin{figure}[tb]
	\centering
	\includegraphics[width=0.495\linewidth]{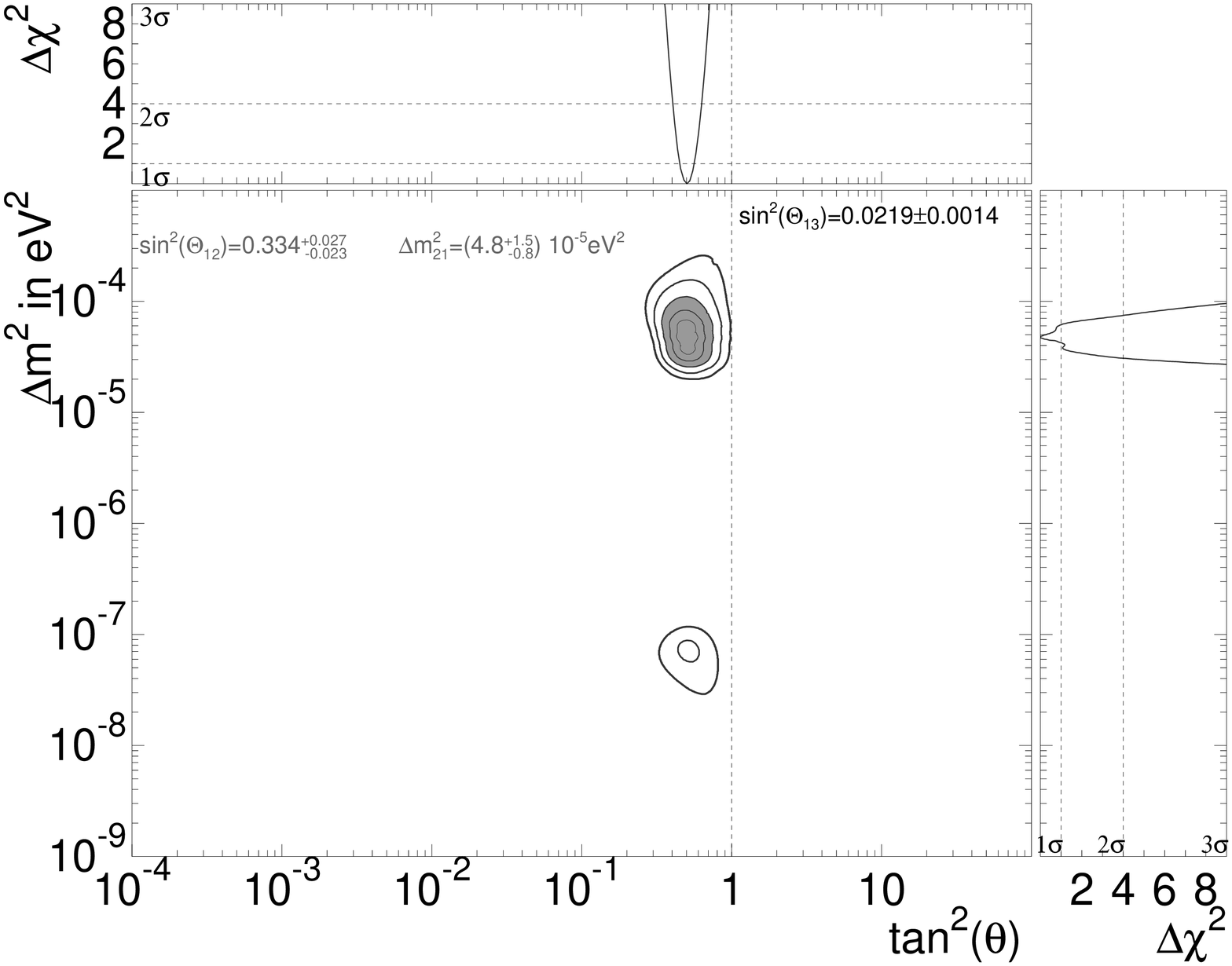}
	\includegraphics[width=0.495\linewidth]{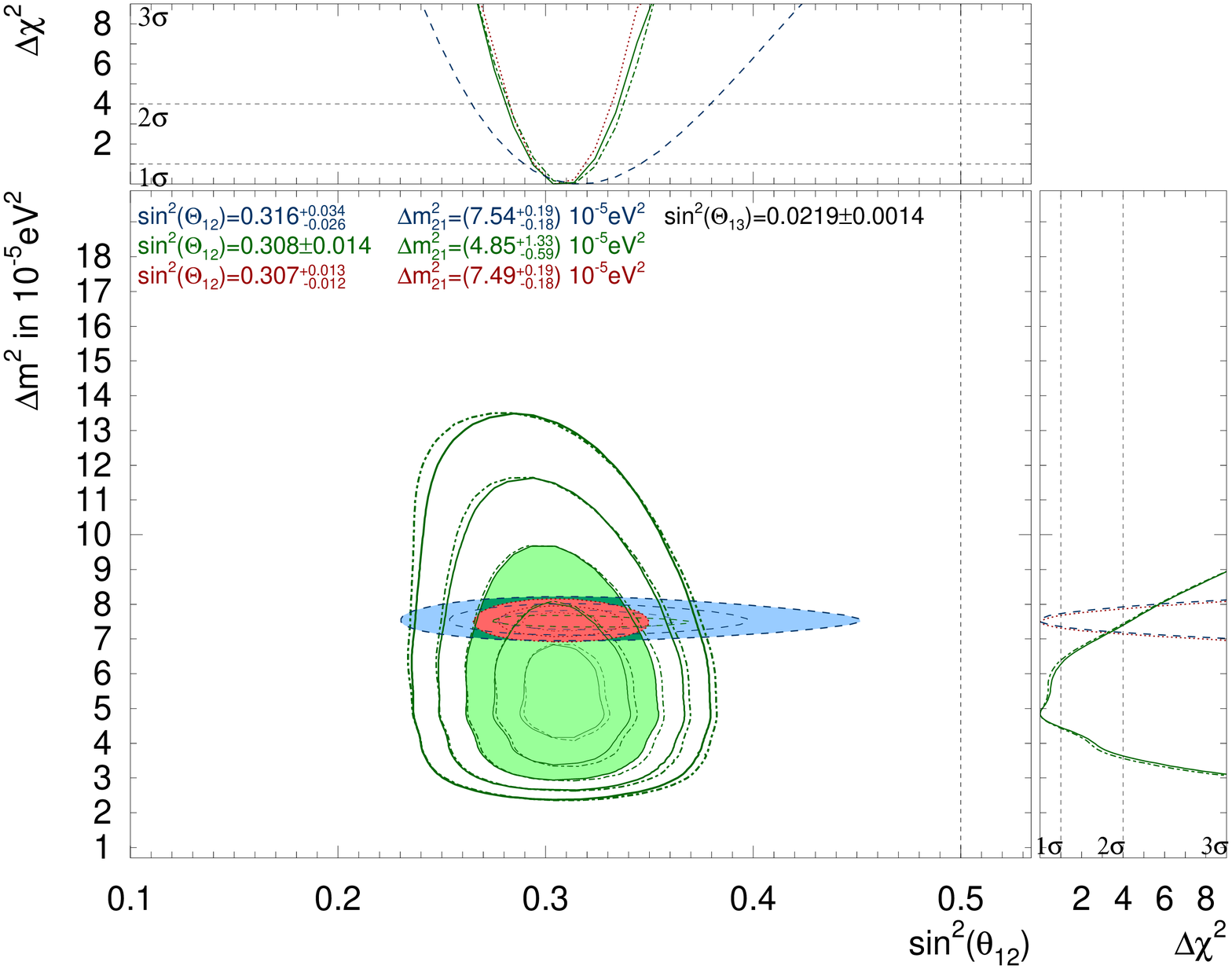}
	\caption{The Super-Kamiokande fit to $\nu_{e}$ oscillations, left.
	Right has the combined fit with SNO and KamLAND data.}
	\label{fig:dnspeccont3}
\end{figure}
%

The Kamiokande detector adopted solar neutrinos as a physics goal shortly
after it was clear that the proton was not going to decay at the predicted
rate.  Kamiokande was smaller and deeper, at 2030 meter water equivalent and so had a muon rate of 0.4 muons
per second.  To accomplish the solar neutrino goal required changes to
the experiment.  Radioactivity such as radon had to be removed from the
water, the energy threshold needed to be lowered, gamma rays from the surrounding rock needed to be reduced and timing was added to
the electronics.  The University of Pennsylvania joined the collaboration
to provide the electronics.  A water purification system was added to remove
radon and an instrumented water shield was added to attenuate gamma rays
from the rock.  The upgraded detector was named Kamiokande-II

First results on solar neutrinos were reported in 1987\cite{KamStat}.
These were based on a fiducial mass of 680 tons, 450 days of observation.
The trigger threshold was 6.7 MeV at 50\% efficiency rising to 90\% at 8.8 MeV.  The analysis threshold was taken above 9.3 MeV recoil electron energy.
Angular resolution at 10 MeV was 28$^{\circ}$ and the energy resolution
at this energy was 22\%.  The observations were ($46\pm13\pm8$)\% of the expected 
value of the $~^{8}B$ neutrino flux of $5.8 \times 10^{6} \nu/cm^{2}/s$.
During a similar period the $~^{37}Cl$ experiment reported
($53\pm20$)\% of the expected signal.  This was a very important result
since it removed doubts about experimental inefficiencies for either
project and shifted attention toward possible problems with neutrino propagation or the solar model.

Kamiokande-II was improved by doubling the phototube gain.  The energy resolution at 10 MeV dropped to 19.5\%.  The trigger threshold was lowered
to 6.1 Mev at 50\% and 7.9 Mev at 90\%.  The analysis threshold was dropped
to 7.5 MeV.  Based on 288 days of data through April 1989 ($39\pm9\pm6$)\% of the expected flux was reported.  These results are summarized in
figures \ref{fig:kamikandesolnu} and \ref{fig:kamikandesolnuE}.

A major motivation for the construction of Super-Kamiokande was a better
measurement of solar neutrinos.  The mass used for solar neutrino observations
was 22.5 ktons, as compared to the 0.68 ktons for Kamiokande.
Super-Kamiokande started with an energy threshold of 4.49 MeV but was eventually able to lower it to 3.49 MeV
The detector started operation in 1996 and went through 4 configurations.
The solar neutrino flux, assuming 100\% electron neutrinos, measured in each of those phases\cite{SKSol} is shown in the table \begin{center}
 \begin{tabular}{|cc|}
	\hline
	Phase & Flux ($\times 10^{6}/(cm^{2}s)$ \\
\hline
	SK-I & 2.380 $\pm$ 0.024 $^{+0.084}_{-0.076}$ \\

	SK-II & 2.41 $\pm$ 0.05 $^{+0.16}_{-0.15}$\\

	SK-III & 2.404$\pm$ 0.039 $\pm$ 0.053\\

	SK-IV & 2.308$\pm$ 0.020 $^{+0.039}_{-0.040}$\\

	Combined & 2.345 $\pm$ 0.014 $\pm$ 0.036\\
	\hline
\end{tabular}
\end{center}
 \begin{figure}[tb]
 	\centering
	\includegraphics[width=0.525\textwidth]{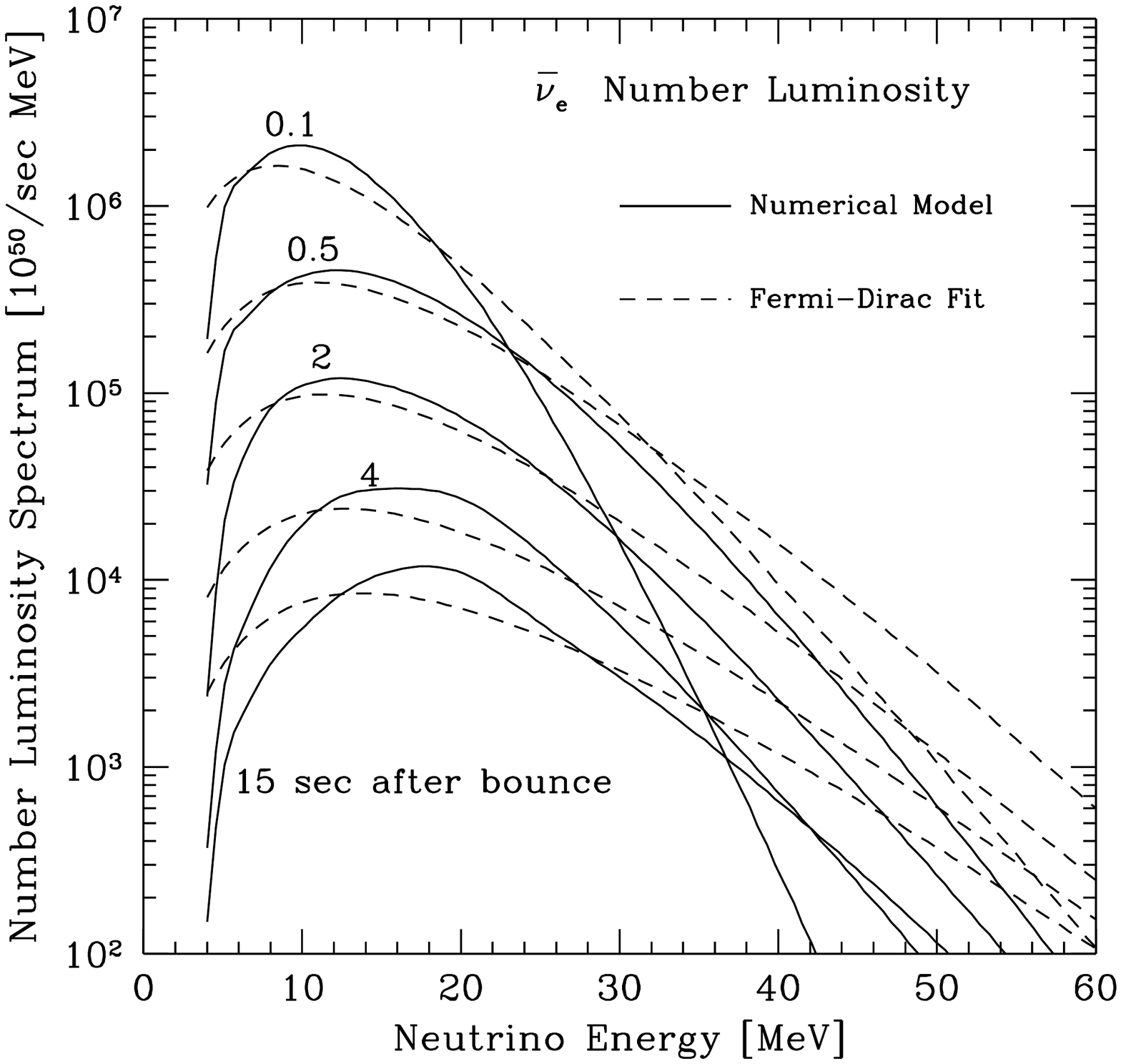}
	\includegraphics[width=0.525\linewidth]{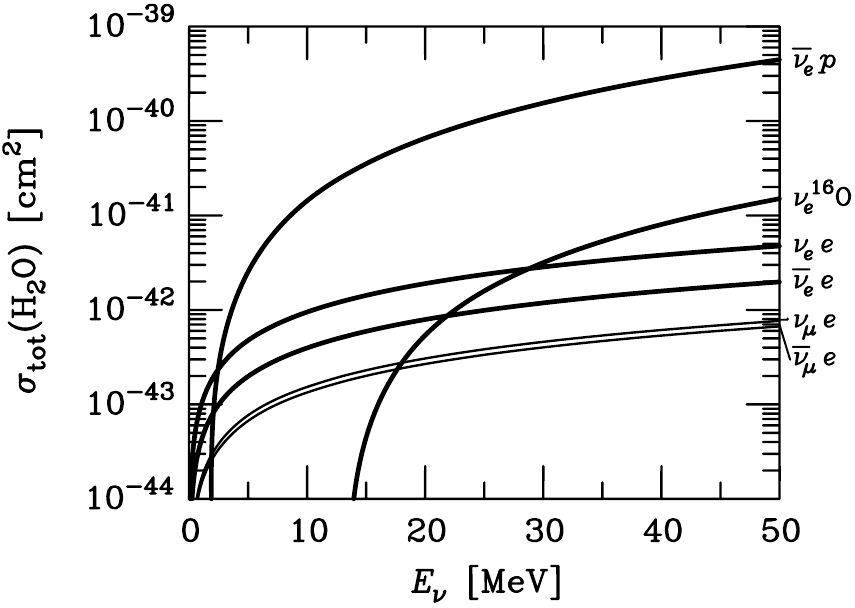}
	\caption{Top: Energy spectra of $\bar{\nu}_{e}$ at different times and the Fermi-Dirac distribution
		with the same average energies \cite{TotaniSatoDalhedWilson98}. The chemical potentials of the Fermi-Dirac distributions are set to zero.
		Bottom: The cross section per molecule for various interactions of neutrinos on water. The plot spans the expected range of supernova neutrino energies. The interaction rate is clearly dominated by $\bar{\nu}_{e} p$ charged current scattering.  Taken from reference \cite{Raf}
		\label{fig:anue_spe}.}
\end{figure}
\begin{figure}[tb]
	\centering
	\includegraphics[width=0.65\textwidth]{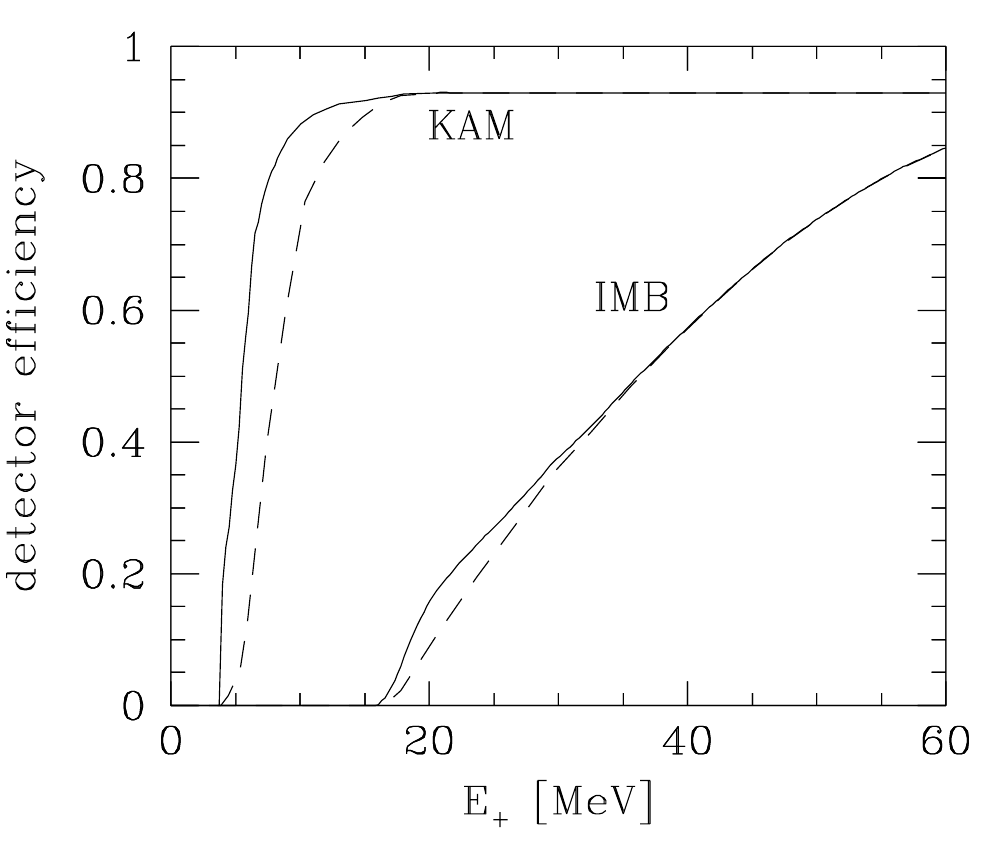}
\caption{The detection efficiency for supernova $\bar{\nu}_{e}$ for
	Kamiokande and IMB.  The solid curves are for the positron
	energy.  The dashed curve is for the neutrino energy.  Taken from
	reference \cite{SNNuO}}
	\label{fig:rafellt2012fig45}
\end{figure}

Since the direction of the neutrino is known it is possible, in principal,
to reconstruct the solar neutrino energy from just the scattering angle and
the energy of the final state electron.  Multiple scattering of the recoil
electron makes this impractical so fits have been done to the electron recoil energy.

Solar neutrino oscillations, which are dominated by MSW effects in the source
also can undergo MSW induced oscillations in passing through the earth.
A two component neutrino oscillation analysis parameterized by
$\Delta m_{12}^{2}$ and $\theta_{12}$ strongly constrains solar
neutrino oscillations, figure \ref{fig:dnspeccont3}.  Super-Kamiokande results combined with those from SNO and KamLAND, clearly removes the vacuum oscillation solution and are also shown in figure \ref{fig:dnspeccont3}.
The results on oscillation parameters were:
$\Delta m_{12}^{2} = 4.8 ^{+1.5}_{-0.8} \times 10^{-5}$ eV$^{2}$
and $\sin^{2}(\theta_{12})=0.334 ^{+0.027}_{-0.023}$

\subsection{Supernova Neutrinos}
A star that has consumed its nuclear fuel can no longer support itself
against gravity.  The star contracts and goes through
many burning stages until it can not support its mass via thermal pressure.
If the star is massive enough there is no other source of support.  The
star collapses under its own weight.  Neutrinos play two roles in the collapse
process.  A neutronization process in the core converts all electrons and
protons to neutrons which removes the electron Fermi gas pressure in the
core and initiates a collapse of the star\cite{HowSN}.  The collapse converts massive
amounts of gravitational binding energy ($3\times 10^{53}$ ergs) to heat.
The heat is trapped by all of the mass above it.  The only thing that can leak out of the star are neutrinos.  Neutrinos in thermal equilibrium
rapidly diffuse and cool the star.  The neutronization process also produces
one electron neutrino for every neutron created.  These neutronization neutrinos can escape before the collapse has made the star opaque to neutrinos.

Supernova neutrino signals from near enough supernovae are observable with
Earth based neutrino detectors.  One expects a transient signal correlated
with optical observations.  The neutrino burst lasts about 10 seconds and
nearby supernovae are rare events.  The estimated rate of supernovae in our galaxy is $4.6^{+7.4}_{-2.7}$ per century\cite{ONGS}.
The last supernova seen in our galaxy
was in 1604 over 400 years ago.  Since the light and neutrinos travel at
(near) the speed of light there is no way to anticipate a supernova
neutrino burst.  Observations must be made with massive detectors capable
of recording neutrinos of modest energies and operating them as much as possible.

Figure \ref{fig:anue_spe} shows an estimated electron antinuetrino spectrum
from a supernova.  The energy stretches from a few MeV to 60 MeV
but is peaked at about 15 to 20 MeV.  The flux drops as a function of time.
Note most of the flux is gone after 15 seconds.  The dashed curves in figure
\ref{fig:anue_spe} are a Fermi-Dirac distribution with the same average energy.  The different neutrino energies have different opacities and so are released at varying equilibrium temperatures, distorting the spectrum.
The rate of neutrino interactions depends on the flux, as in figure \ref{fig:anue_spe} times the cross section as in figure \ref{fig:reactorxsec} for $\bar{\nu}_{e}$ charged current reactions on hydrogen. 

Observation of neutrinos from a supernova burst is usually done through
observations of the $\nu_{e}$ and $\bar{\nu}_{e}$ charged current reactions.
While all neutrino types are produced and neutrino oscillations mix all the
neutrino types together neutrinos of other types are below the charged
current threshold so can only interact via neutral currents.  The neutral
current has a smaller cross section and gives only a fraction of the neutrino energy deposition.  Neutral currents provide a less useful, more ambiguous, signal for observations. 

\begin{table}[ht]
\begin{center}
	\begin{tabular}{|cccc|cccc|} \hline
		Kamiokande: 1987 & & & & IMB: 1987 & & &  \\ \hline
		event & time & energy        & angle        & event & time & energy  & angle    
		\\
		& (sec)  & (MeV)          & (deg)        &   & (sec) & (MeV)      & (deg)    
		\\
		1  & 0.000  & 20.0 $\pm$ 2.9 & 18  $\pm$ 18 & 1 & 0.000 & 38 $\pm$ 7 & 80  $\pm$
		10  \\
		2  & 0.107  & 13.5 $\pm$ 3.2 & 40  $\pm$ 27 & 2 & 0.412 & 37 $\pm$ 7 & 44  $\pm$
		15  \\
		3  & 0.303  & 7.5  $\pm$ 2.0 & 108 $\pm$ 32 & 3 & 0.650 & 28 $\pm$ 6 & 56  $\pm$
		20  \\
		4  & 0.324  & 9.2  $\pm$ 2.7 & 70  $\pm$ 30 & 4 & 1.141 & 39 $\pm$ 7 & 65  $\pm$
		20  \\
		5  & 0.507  & 12.8 $\pm$ 2.9 & 135 $\pm$ 23 & 5 & 1.562 & 36 $\pm$ 9 & 33  $\pm$
		15  \\
		6  & 1.541  & 35.4 $\pm$ 8.0 & 32  $\pm$ 16 & 6 & 2.684 & 36 $\pm$ 6 & 52  $\pm$
		10  \\
		7  & 1.728  & 21.0 $\pm$ 4.2 & 30  $\pm$ 18 & 7 & 5.010 & 19 $\pm$ 5 & 42  $\pm$
		20  \\
		8  & 1.915  & 19.8 $\pm$ 3.2 & 38  $\pm$ 22 & 8 & 5.582 & 22 $\pm$ 5 & 104 $\pm$
 20  \\
9  & 9.219  & 8.6  $\pm$ 2.7 & 122 $\pm$ 30 &   &       &            &          
\\
10 & 10.433 & 13.0 $\pm$ 2.6 & 49  $\pm$ 26 &   &       &            &          
\\
11 & 12.439 & 8.9  $\pm$ 1.9 & 91  $\pm$ 39 &   &       &            &          
\\
\hline
\end{tabular}
\end{center}
Table from \cite{KST} \label{tab:SN1987A}
\end{table}
\begin{figure}[tb]
	\centering
	\includegraphics[width=0.5\linewidth]{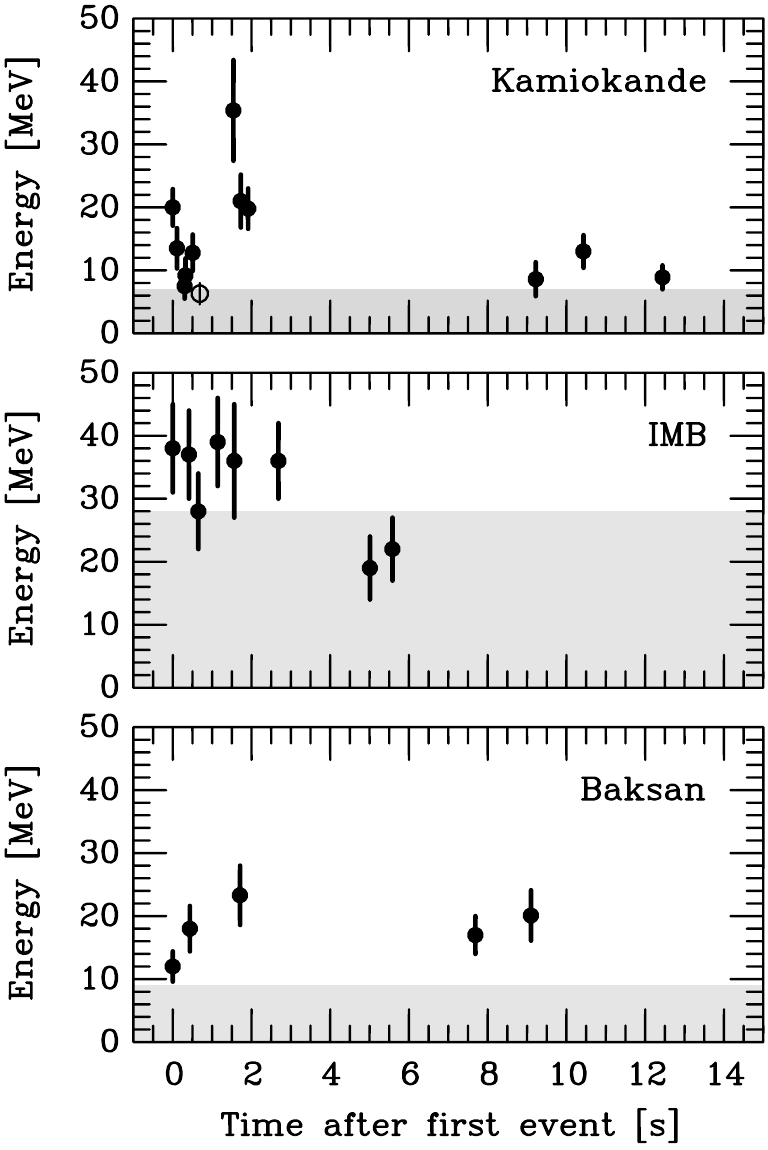}
	\caption{A plot of the energy and time data from table \ref{tab:SN1987A}
	The gray region shows the energy region below 30\% detection efficiency.
Events from the Baksan detector, composed of liquid scintillator are also shown.  The time scales are synchronized on the first event.  Figure from reference \cite{Raf}}
	\label{fig:rafellt2012fig47}
\end{figure}

On February 23, 1987 a neutrino burst originating in the LMC galaxy was observed by
Kamiokande, IMB and a liquid scintillation detector in Baksan USSR.
The source of the neutrinos was associated with a visible supernova, confirming astrophysical models of the role of neutrinos in stellar collapse.
The distance to the source was outside our galaxy, about 160,000 light years
away.  The eight events reported by IMB and the eleven events reported by
Kamiokande are listed in table \ref{tab:SN1987A}.  The observable quantities
were the arrival time, the energy of the recoiling positron and the angle
of the positron with respect to the neutrino direction.  The times and energies are plotted in figure \ref{fig:rafellt2012fig47}.  To convert these observations into useful physics results needs an understanding of the
detection efficiency, figure \ref{fig:rafellt2012fig45} and the fiducial volume, the target mass.

The times in table \ref{tab:SN1987A} are all relative to the first event
from the burst in that detector.  There was no synchronization between the
experiments.  The IMB time was taken from the WWVB time standard accurate
to about $\pm50$ ms.  The first IMB event was observed on February 23 at 7:35:41.37 UT.
\begin{figure}[tb]
	\centering
	\includegraphics[width=0.9\linewidth]{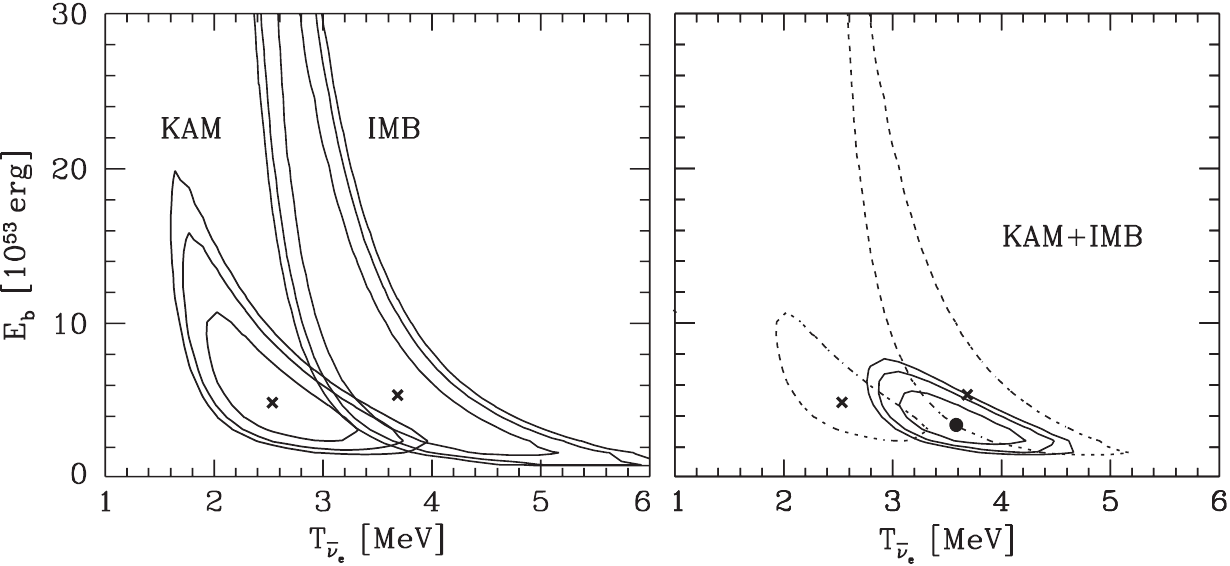}
	\caption{A fit of the total energy in neutrinos, the neutron star binding energy, $E_{b}$ and the electron antineutrino temperature $T_{\bar{\nu}_{e}}$ for both IMB and Kamiokande data and for the combined sample.}
	\label{fig:rafellt2012fig48}
\end{figure}

Analysis of the 1987 supernova data was encouraging at the time.  But neutrino
oscillations were not yet fully understood.  Since supernovae produce all kinds of neutrinos, and they may have different energy distributions, analysis
is complicated by neutrino flavor oscillations effects at the source, in propagation and at detection\cite{SNNuO}.  An analysis of the Kamiokande and IMB supernova observations is illustrated in figure \ref{fig:rafellt2012fig48}.  The fit\cite{Raf} is in fairly good agreement with the expectations of $3\times10^{53}$ ergs and $3.5$ MeV.
\subsection{Point Sources}
The presence of cosmic ray photons above about 1 GeV make it clear that
there are energetic astrophysical processes in the universe that are capable
of producing unstable elementary particles that decay into neutrinos.
Since it is much easier to absorb energetic gamma ray photons than neutrinos
one expects the neutrino flux to be higher than the gamma ray flux.
IMB searched for localized directional sources of neutrinos\cite{Svob1987}
based on a sample of 172 upward going muons.  The use of upward muons restricted observations to when the source was below the horizon.
A search of 10 preselected sources showed no significant excesses.

A search using contained events\cite{IMBPS} gave access to the entire sky but also yielded no excess above background for any of 15 preselected sources.

Super-K has also looked for point sources\cite{Thane} using 2623 days of data
collected from 1996 to 2007.  The search was conducted with upward going muons
coming from neutrino interactions below the detector.  Four searches were conducted, an unbiased search over the whole sky, a search of 16 preselected  astrophysical candidates, a search correlated with 27 ultra high energy cosmic rays from Auger and a search for correlations with about 2200 gamma
ray bursts in the BATSE, HETE and Swift catalogs.  No significant signals were found.

Super-K searches for specific sources\cite{SKBlaz} have also not yielded
any significant signal.
\subsubsection{MACRO}
MACRO has reported on searches for astrophysical neutrinos.\cite{MACRO}.
The data was recorded from March 1989 to September 1999 with a total live time
of about 6.4 years, with varying acceptances.  1100 upward-going muon events
were recorded.
\begin{figure}
	\centering
	\includegraphics[width=0.5\linewidth]{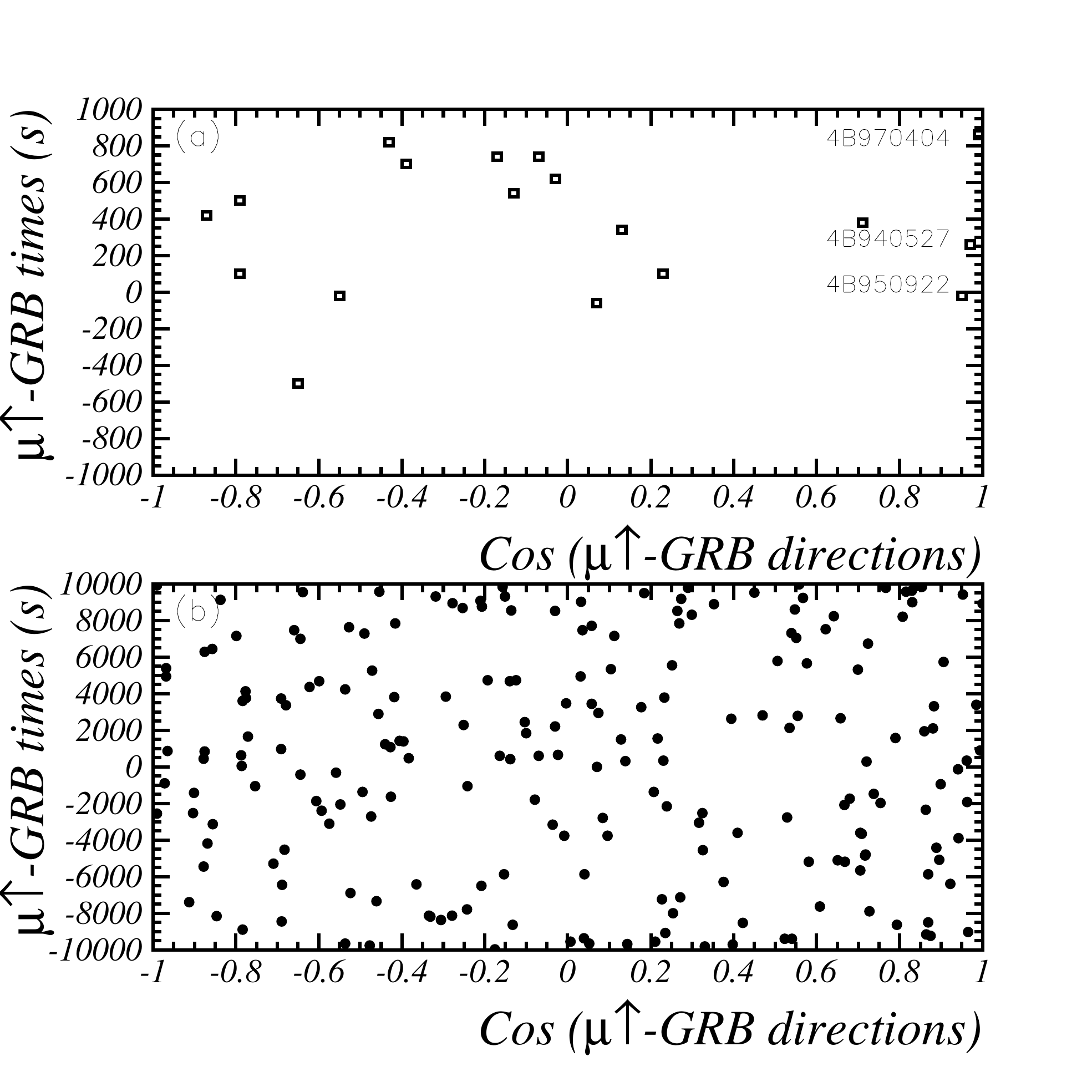}
	\caption{The time and angular differences between the MACRO 1100 upward muon sample and the 2527 BATSE gamma ray burst sample. The lower plot has a time limit of $\pm$10,000 seconds. The upper plot selects the sample within $\pm$1,000 seconds. Three BATSE bursts are labled.}
	\label{fig:macro-batsecorr}
\end{figure}

A search for clusters of upward muon events in a cone of 1.5$^\circ$,
3$^\circ$ or 5$^\circ$ (half angle) of a direction found
60 clusters when 56.3 would be expected from the atmospheric neutrino
background.  None of the clusters were statistically significant.

A search in the direction of 42 potential energetic neutrino sources selected
from known energetic photon point-sources obtained a flux limit of
$\Phi_{lim}(90\%) = 3.06 \times 10^{-16}$ cm$^{-2}$ s$^{-1}$.

They conducted similar searches in the direction of 220 supernova remnants,
181 blazars, 2527 BATSE sources, 271 EGRET sources, 32 BeppoSAX sources
and 29 NovaeX sources with varying limits with none above  
$6.84 \times 10^{-16}$ cm$^{-2}$ s$^{-1}$.

They also looked for time correlations with transient events from
BATSE, figure \ref{fig:macro-batsecorr}
and BeppoSAX.  Results were compatible with atmospheric
backgrounds.

\subsection{Dark Matter}
Strong evidence for dark matter comes from observation of gravitationaly bound
systems of stars and galaxies and of the cosmic microwave background and the large scale structure (the distribution of galaxies) of the universe. \cite{DM}  Conventional understanding of physics suggests that the energy
density observed as dark matter has a massive particle like quanta with
all of the symmetries expected by quantum field theory.  Stable dark matter must have no electric charge or we would find evidence for it in bound states, like atoms, interacting with electromagnetic radiation.  If it had
electric charge it would be visible, emitting and absorbing light.
Dark matter accounts for about 85\% of the matter in the universe, with an energy density of about $2.2\times 10^{-27}$ kg/meter$^3$.  Since it is responsible for most of the gravitational structure in the universe its distribution is not expected to be uniform.

Observation of the particle manifestation of dark matter would be a major step
in understanding its physical nature.  It must be present in most regions
dominated by gravitational binding.  Since it is present in our laboratories
substantial effort has gone into {\em direct detection} where one observes
an interaction between dark matter particles and a conventional detector.
Direct detection is one of three major observational methods that are being
pursued.

Production methods of dark matter detection  use the enormous energies of high energy colliders to produce dark matter out of pure energy.    Production methods
have been used to very effectively fill in almost all of our knowledge about
elementary particles since almost all of them are unstable and have decayed
and are no longer present in our environment.  The idea of production methods
is to create matter and filter out all the components of ordinary matter.
What is left would be a dark matter candidate.

Indirect methods to observe dark matter take advantage of its ability to clump
and the general properties of matter described by relativistic quantum field
theories.  In particular one expects dark matter to come in matter and
antimatter versions.  When matter and antimatter meet they convert to pure
energy.  Even if the dark matter has virtually no particle like interactions
with ordinary matter the energy of annihilation will be detectable.

\subsubsection{Annihilation}
Indirect methods provide a link to neutrino astrophysics.  Annihilation rates
depend on the square of the density of dark matter particles.  Gravitationaly induced clumping of dark matter enhances the annihilation rate.  Because dark matter interacts only weakly with normal matter once the dark matter becomes gravitationaly bound it will accumulate inside of
celestial object.  access to the normal matter annihilation products will be blocked from view except for annihilation products which decay promptly
to neutrinos.  So bounds on energetic neutrinos from the sun and the earth
can be used to discover dark matter.

IMB was the first to look for evidence of dark matter annihilation\cite{DM87}.
Both contained events and through-going muons were studied.  No statistically significant excesses were reported.  Model dependent bounds were placed on
some forms of dark matter.

Super-Kamiokande had multiple searches for evidence of dark matter\cite{SKDM}.
They looked for evidence in upward going through-going muons from the direction of the earth center, the sun and the galactic center.  No significant excess was found.  Limits on the WIMP dark matter proton cross section in the range of $10^{-41} cm^{2}$ to $10^{-38} cm^{2}$ was set depending on the WIMP mass and the assumed form of the coupling.  A latter search with a much larger exposure confirmed these limits.

\subsubsection{Direct Dark Mater Searches}
Many forms of dark matter are expected to be unstable and decay.  Most dark matter searches look for the stable, neutral, lightest particle of the dark matter class.  There is no reason why unstable dark matter with a very long lifetime would not persist to the present time, like the naturally occurring
thorium 232, uranium 238 and potassium 40.  The ambient dark matter would then present an unshieldable component in high sensitivity experiments
such as proton decay.  The signal would not look like proton decay
unless there was an unlikely coincidence of mass scales.

Unlike the atmospheric neutrino component the decay of dark matter should have
no net momentum but this constraint can be confused by decays to lighter
dark matter particles which would carry off momentum unseen.  So the decay
of long lived metastable dark matter would appear as an isotropic anomalous
component of the atmospheric neutrino signal.  Such a ``component'' was
found\cite{NuAnom}
but was eventually identified as a consequence of neutrino oscillations.

A Super-Kamiokande did a direct search for boosted dark matter coming from the sun or
the galactic center with a 161.9 kt yr exposure.  The search yielded no significant signal.  Boosted dark matter is a form of dark matter which is energetic and may have been produced from cold dark matter.  The search was for energetic electrons recoiling from the dark matter
as it crossed the detector.  They looked for recoiling electrons in the energy
range of $0.1<E_{e}<1000$ GeV.  Signal rates closely matched estimates due to atmospheric neutrinos.
\subsection{Magnetic Monopoles}
In addition to proton decay grand unified theories predicted the existence of
magnetic monopoles\cite{HoftPoly}.  A magnetic monopole is an object with a single magnetic
charge.  All known magnetic states are composed of magnetic dipoles containing
both north and south magnetic charges.  The existence of stable magnetic monopoles meant that they would be produced during the Big Bang and would survive until the present day.  The nature of grand unified theories also
gave these monopoles the power to catalyze proton decay.  A grand unified
magnetic monopole would cause the baryon violating decay of protons and
neutrons in which it came into contact\cite{Rubakov}.

The IMB searches for monopole catalysis of proton decay\cite{SE1983} yield a flux limit
of from $1.0 \times 10^{-15}/(cm^{2} sr sec)$ to $2.7 \times 10^{-15}/(cm^{2} sr sec)$ at 90\% confident limit.

The Super Kamiokande search for magnetic monopoles\cite{SKMon} used a different method.  They looked for evidence for monopole catalyzed proton decay in the sun which would produce pions which would ultimately decay
to neutrinos with energies in the range of 19 to 55 MeV.  By searching for
a flux of these energetic neutrinos from the sun they set a limit on the
monopole flux of $F_{M}(\sigma_{0}/1mb)<6.3 \times 10^{-24} (\beta_{M}/10^{-3}) /(cm^{2} sr sec)$ at 90\% confidence limit.
$\sigma_{0}$ is the catalysis cross section at $\beta_{M}= 1$ and $\beta_{M}$
is the monopole speed in units of $c$.

MACRO was able to set direct limits on magnetic monopole flux
by measuring the ionization and transit time of possible candidates.  No events were found setting 90\% confidence limits at $< 2 \time 10^{-16}/(cm^{2} sr sec)$ over the range
of velocities $ 4\times 10^{-5}<\beta<1$.  They also set limits
on slow moving GUT monopole catalyzed proton decay of
  $< 3 \times 10^{-16}/(cm^{2} sr sec)$
\subsection{Muons}
Muons present an opportunity to significantly extend the sensitivity of
astrophysical neutrino searches.  Muons are an unstable elementary particle
with $c\tau=659$ meters and a mass of 106 MeV.  Virtually all muons observed on earth were created nearby.  Muons can be created by the decay of heavier
elementary particles, primarily pions or much less likely by neutrino
interactions.  The muons observed in cosmic rays come primarily from the
decay of a charged pion into a muon and a neutrino.  The pions are produced
via strong interactions of cosmic rays on the atmosphere.

Whenever a muon is produced a neutrino is produced so cosmic rays produce
neutrinos in the Earth's atmosphere.  These neutrinos can be observed when they interact in the Earth and produce a muon.  The first evidence of
atmospheric neutrinos was the observation of horizontal and upward going
muons in deep mines where all surface muons had been attenuated.\cite{atmu}

Many astrophysical neutrino observatories extend the range of their observations to much lower fluxes, at higher energies by looking at energetic
neutrinos emerging from the surrounding rock.  Some information is lost.
The interaction vertex can not be observed so energy deposited near the
interaction point is lost.  The range of the muon in the rock is also not known so the target mass can only be estimated with a model of the interaction
spectrum.

The muon is a charged particle that has no strong interaction and a relatively
long lifetime.  Time dilation gives high energy muons a much larger lifetime
in our reference frame.  Because of these two factors muons are the dominant
source of cosmic rays found in underground laboratories.  Except for radioactive decay of materials in the underground environment they are the
dominant source of radiation underground.  Several experiments have attempted
to extract evidence for an astrophysical source from downward going muons
in underground experiments.  These muons can be associated with the direction
and timing of known high energy astrophysical sources such as AGN.  Since the muon is electrically charged and would be deflected by magnetic fields en route and the muon is unstable so would decay en route muons could be associated with the source if they originated from the interaction of a long lived neutral particle coming from the source.

IMB\cite{IMBCygX3} searched for a signal from the direction of Cygnus X3
which is a known, variable source, of periodic energetic high energy
gamma rays.  Searches were made at times of reported activity from both
surface and underground detectors.  No significant results were observed.
Kamiokande\cite{KamCygX3} has reported similar results.

Super-Kamiokande\cite{SKCR} has shown evidence for anisotropy in the high energy cosmic ray flux by measuring the flux of downward going muons from a
large portion of the sky.
\section{Future Opportunities}
\subsection{Hyper K}
Hyper-Kamiokande  consists  of  a  cylindrical  tank of 260,000 tons of water,  with  a  height  of  60m  and  a  diameter  of  74m.  surrounded by 40,000 phototubes at a depth of 650 mwe.  The goals of Hyper-Kamiokande
are similar to those for Supper-Kamiokande.  The additional mass will give it
more events for solar, atmospheric and supernova neutrinos.  The additional mass will also increase the event rate for the accelerator neutrino beam
giving it access to the neutrino mass hierarchy and neutrino CP violation.
\subsection{JUNO}
JUNO consists of 20,000 metric tons of liquid scintillator observed by
15,000 phototubes.  The experiment plans to use reactor antineutrino proton scattering
to resolve the neutrino mass hierarchy question but will also have a strong
astrophysics program.  They plan to have an energy resolution of better
than 3\% and employ a delayed coincidence method to suppress backgrounds.
\subsection{DUNE}
DUNE plans to study neutrinos interacting on 40,000 metric tons of liquid argon in four 10,000 ton modules.  One of the primary goals of the experiment is to make observations
of the electron neutrino component (as distinct from the electron antineutrino component) of a supernova neutrino burst\cite{dunesn}.  Argon has a relatively
large cross section for electron neutrinos.  The neutronization process
that forces electrons into protons to make a neutron star is a key part
of initiating the stellar collapse mechanism and it produces an electron neutrino pulse preceding the much larger pulse of thermal neutrinos powered by the neutron star binding energy.  Observation of neutrinos in DUNE should be complementary to other detector's observations of the antineutrino
content via the  antineutrino proton scattering mechanism
\section{Conclusions}
Neutrino astrophysics is a thriving subject with many established discoveries.
Many neutrino mysteries have been solved by massive detectors which discovered
and measured neutrino oscillations which enabled correct interpretations of astrophysical observations.  Oscillations among three neutrino flavors complicate neutrino astrophysics since the three flavors produce different signals.  One needs to work backward from the observations to understand the source.  This worked well for the sun.
\section{Acknowlegements}
I would like to thank John G.~Learned for suggesting this
article and helping me expand it.  I thank Floyd W.~Stecker
for his organization and helpful suggestions to improve the
manuscript.  This article would not have been possible if
not for the contributions of those physicists and
astronomers who have created this field.  The citations
can not do all of them justice.  Finally I owe a debt of gratitude to all of my collaborators over the years who have
patiently helped me learn and understand the subject.

\end{document}